\documentclass[pra,preprint,showpacs]{revtex4-2}
\usepackage[centertags]{amsmath}
\usepackage[utf8]{inputenc}
\usepackage{amsfonts}
\usepackage{amssymb}
\usepackage{amsthm}
\usepackage{amsmath}
\usepackage{newlfont}
\usepackage{epsfig}
\usepackage{amscd}
\usepackage{graphicx}
\usepackage{epsfig}
\usepackage{footnote}
\usepackage{lipsum}
\usepackage{color}
\usepackage{xcolor}
\usepackage{graphicx}
\usepackage{multirow}
\usepackage{enumerate}
\usepackage{caption}
\usepackage{subcaption}

\usepackage{amscd}

\newcommand{\beq}{\begin{equation}}
\newcommand{\eeq}{\end{equation}}
\newcommand{\ba}{\begin{array}}
\newcommand{\ea}{\end{array}}
\newcommand{\bea}{\begin{eqnarray}}
\newcommand{\eea}{\end{eqnarray}}
\begin{document}

\begin{center}
{\large \sc \bf  {{   Probabilistic} perfect pure quantum state transfer via state  restoring and ancilla measurement}
}

\vskip 15pt

{\large
 E.B.~Fel'dman$^{1}$,  J.Wu$^2$, A.I.~Zenchuk$^{1,*}$
}

\vskip 8pt

{\it $^1$
Federal Research Center of Problems of Chemical Physics and Medicinal Chemistry RAS,
Chernogolovka, Moscow reg., 142432, Russia}.

{\it $^2$ School of Mathematical Sciences, Zhejiang University, Hangzhou 310027, PR~China}

{\it $^*$Corresponding author. E-mail:  zenchuk@itp.ac.ru}

\vskip 8pt

\end{center}


\begin{abstract}
We  propose  the protocol for {   the probabilistic} perfect  state transfer of an arbitrary {multiqubit} pure quantum state  along the spin-1/2 chain governed by the Hamiltonian preserving the excitation number in the system.  We show that the  {multiqubit}  $k$-excitation pure sender's state can be restored at the receiver using only the local transformations  over the qubits of the extended receiver. The restored state appears in the superposition with other states which form garbage. This garbage can be easily removed by including the ancilla, whose state  labels the garbage, and then  measuring the  {ancilla state} with desired output.  The resulting state of the receiver coincides with the initial sender's state.   Then, to transfer an arbitrary multiqubit pure state of some system $S^{(0)}$,  we encode this state   into the $k$-excitation state of the sender, transfer and restore this state and finally decode  the restored $k$-excitation state {of the receiver} into the  state of another subsystem $R^{(0)}$.  After labeling and removing the garbage via the ancilla-state measuring {   with certain success probability}  we complete the algorithm for the {    probabilistic}  perfect transfer of an arbitrary pure state.
\end{abstract}


\maketitle

\section{Introduction}
The quantum state transfer is a very popular problem in quantum informatics. First, it was formulated by Bose in his famous paper \cite{Bose}. Of course, the most  attractive idea was to realize the perfect state transfer (PST), in which case the state of the receiver at some time instant coincides with the  initial sender state. Set of papers were devoted to such a problem, among the first ones we refer to \cite{CDEL,KS}, where this aim was achieved through the special selection of the coupling constants in the nearest-neighbor $XX$-Hamiltonian. Since the PST requires very precise adjusting of coupling constants, it  is very sensitive to their  perturbations. The stability problem was also intensively studied \cite{ZASO,ZASO2} and it  was demonstrated that the  
high-probability state transfer that can be organized by suitable adjustment of the end-node coupling constants
\cite{VBR,VGIZ,GKMT,GMT} is more stable to perturbations and much simpler for realization. 

 Nevertheless, the PST in various quantum communication lines  attracts attention of many researchers. Usually, this phenomenon requires special geometry and very precise choice of parameters of the interaction Hamiltonian, such as coupling constants and selected directions of spin-spin interactions. There are many graph models simulating the PST  where vertices are associated with spins (qubits) and edges correspond to the specific interactions between appropriate spins. Thus, the PST between  vertices of a graph in the  special discrete-time quantum walk model called Grover walk is studied in 
 \cite{KS_2022}.  In  \cite{D_2022}, the PST between any two vertices is organized  in Markovian quantum walk. The PST between distinct atoms in generalized honeycomb nanotori is considered in \cite{AAJZ}. 
The method for constructing 1D Hamiltonians having the structure of a Jacobi matrix and realizing the PST is proposed in  \cite{BCDFZ} and  based on the properties of the Krawtchouk and Chebyshev polynomials. 
Some other examples relating the PST with Jacobi matrix and orthogonal polynomials are given  in \cite{B_2024}.  The review  of protocols for  the PST and Pretty
Good State Transfer based on the graph theory is presented in \cite{KGA,AK}.  Disadvantage of such methods is their high sensitivity to perturbations of coupling constants that destroy PST reducing it to the high-probability state transfer. 

The High-fidelity state transfer (HFST) is also a very popular direction in development of quantum information transfer. 
Among other methods we select the method based on weak end-bonds in a spin chain. First, this method  was introduced in Ref.\cite{VBR} for studying the  non-ideal teleportation via a spin-chain and in Ref.\cite{VGIZ} for 
optimizing the  long-distance  entanglement.  Then, the chain with two empty sites next to the end-sites was used in   Refs. \cite{GKMT,GMT} for organizing nearly-perfect  state transfer. However,   very long time interval was required 
for realizing such transfer. The problem of long time-interval was  smoothed in Refs.\cite{ZASO,ZASO2} where, along with the stability problem, the method of  optimized end-bonds have been considered and  optimization 
was maneuvering between the fidelity and time-interval  of state transfer. The  weak bonds  properly distributed along the  spin chain were used in Ref.\cite{FZ_2009} for high fidelity state transfer among  various sites of a spin chain. 
As an alternative to the weak-bond method, the inhomogeneous magnetic field can be implemented. In this case, the high-fidelity state transfer can be organized in the homogeneous spin chain by adjusting the Larmor frequencies at the  end-nodes of the spin chain  \cite{DZ_2010}. 
Recently, the optimization of state-transfer fidelity via weak bonds in disordered Heisenberg model was 
proposed in Ref.\cite{LMDMBA}. 
Optimization of entanglement transfer by adjusting weak end-bonds was studied in Ref.\cite{BACVV}.
Two pairs of properly adjusted weak end-bonds \cite{ABCVV} can significantly increase the fidelity of state transfer decreasing, simultaneously, the state-transfer time-interval. The fidelity controlled by the weak bonds   (not necessary end-bonds) in  multi-dimensional spin arrays  governed by the XXZ-Hamiltonian  is considered in \cite{HAPM}.
The HFST driven by inhomogeneous magnetic field was also considered in Refs.\cite{KLYP,LSU,SCT}.

Nowdays, state transfer under interaction with environment attracts attention of many researchers.
Thus, the high-fidelity state transfer along the spin-chain governed by the nearest-neighbor XX-Hamiltonian 
 with properly adjusted coupling constants  and  interacting with environment via the Lindblad operators is studied in  Ref.\cite{XALW}.  State transfer along a short homogeneous chain with non-Markovian dynamics is considered in Ref. \cite{WRLYW}.  The  high-fidelity adiabatic state transfer (in particular, including interaction with environment) is studied in \cite{VBRR}.

Numerous papers were devoted to modification of the state transfer algorithm. Among others, we  select  the remote state creation \cite{XLYG,PBGWK,PBGWK2,DLMRKBPVZBW,PSB,LH,Z_2014,BZ_2015} that is practically realized for the photon systems  \cite{PBGWK,PBGWK2,DLMRKBPVZBW}. In this case, there is a map between the parameters of the sender state and the parameters of the receiver state. However, the region  of creatable parameters does not cover the whole state space of the receiver.

Recently the state restoring of the mixed state was proposed as a method for obtaining the receiver state $\rho^{(R)}$  that has the most close structure to the initial sender state $\rho^{(S)}(0)$ \cite{FZ_2017,FPZ_2021}. 
{Namely, we call the receiver state $\rho^{(R)}$ restored if, at some time instant, each its element is proportional to the appropriate element of the initial sender's states $\rho^{(S)}(0)$, i.e.,
\begin{eqnarray}\label{r1}
\rho^{(R)}_{ij} = \lambda_{ij} \rho^{(S)}_{ij}(0),\;\; |\lambda_{ij}|\le1.
\end{eqnarray}
Here, the coefficients $\lambda_{ij}$ {(complex in general)} are called the $\lambda$-parameters. They do not depend on a particular state $\rho^{(S)}(0)$ to be transferred and are  defined by the interaction Hamiltonian and time instant for state registration. It is important, that the proposed state-restoring algorithm  is not sensitive to a particular Hamiltonian with the only requirement that this Hamiltonian  conserves the excitation number in a system.
We emphasize that, in general, the restoring formula (\ref{r1}) is applicable to the non-diagonal  elements of  $\rho^{(R)}$, while the diagonal elements  require special treatment \cite{FPZ_2021,BFLP_2022}. 
Not all of them can be restored unless the special restrictions on the structure of $\rho^{(S)}$ is imposed. }
The state restoring  is  achievable by introducing special  unitary transformation at the so-called extended receiver (receiver with several closest  nodes of the transmission line). 

{  In this  paper, we present another insight into the problem of PST, {   which we refer to as the probabilistic PST in contrast to  the  deterministic PST discussed above}. 
 {   The probabilistic PST  is the first  practically meaningful algorithm 
 implementing the  state-restoring algorithm as a subroutine.}  
{First, we slightly modify the  state-restoring algorithm  for  the purpose of   pure state restoring.  In this case the restored pure state is in the superposition with other states which form the garbage to be removed later.  In addition, we require that all $\lambda$-parameters equal to each other.  Thus, there is only one parameter $\lambda$.  Then we involve the special ancilla to select the above restored  part (proportional to $\lambda$) from the superposition state.  Finally, we  measure the state  of the ancilla  {    with certain probability of the desired result (the success probability)} to remove all the garbage and keep only the restored part. Due to the normalization, the parameter $\lambda$ disappears from the final state  thus providing the PST.}
In this way we  show that restoring the transferred state with all equal $\lambda$-parameters, $|\lambda|<1$,  supplemented  by the measurement  of the ancilla state allows to establish the {   probabilistic} perfect  transfer of an arbitrary  state. } However, the privilege is given to the so-called  $k$-excitation state (the superposition of basis states with $k$-excitations only) because these states can be {   probabilistically} perfectly transferred involving only local transformations at the receiver site. Then, the {   probabilistic} PST of an arbitrary {multiqubit} pure state can be performed encoding this  state into the $k$-excitation state of the sender,  organizing  the PST of this $k$-excitation state  and then decoding the perfectly transferred   $k$-excitation state  of the receiver into the original pure state. The main obstacle here is the  probability of measuring the required ancilla state {  (success probability)}, which equals $|\lambda|^2$ and can be small in general.  Therefore the optimization of the restoring algorithm to get the largest possible absolute value of  the  $\lambda$-parameter is a relevant problem. { An example of such optimization via {   adjusting}  the weak  end-bonds is considered below  for the 42-spin chain, see Table \ref{Table:T2}.}
 
 {  
 The algorithm proposed  in our paper reveals the following features and advantages.
 \begin{enumerate}
 \item
At most the  restricted number of coupling constants is under control. These constants serve to optimize the spin chain for providing the high-fidelity state transfer. Other coupling constants can be arbitrary. Their values effect on the parameters of the restoring unitary transformation applied to the extended receiver but can not destroy the realizability of {   probabilistic} PST.  
\item
{The restoring matrix applied to the extended receiver can be found for any set of coupling constants in the Hamiltonian. Therefore, the perturbations of coupling constants require only deformation of the restoring unitary transformation  to preserve the state-restoring and thus performing the {   probabilistic} PST.  Any perturbations are not principal for our algorithm provided that they do not destroy the excitation-preserving dynamics {   and do not depend on time}. This is an important advantage in comparison with the algorithms of the {   deterministic}  PST  proposed in the above quoted references which are very sensitive  to perturbations of coupling constants.}
\item
{   The proposed PST is a probabilistic process. The probability of the desired ancilla-state registration at some time instant $t_0$  is defined by 
$|\lambda|^2$. However, once the desired ancilla-state is registered, the associated receiver's  state coincides  with the initial sender's state. In other words, the fidelity of the transferred state is always 1 and the PST is guaranteed.}
\item
{    For the fixed dimensionality of the extended receiver, there are two method of increasing the probability of  PST. The first one (mentioned above in n. 1)  is based on organizing the high-fidelity state transfer using any of the methods proposed in the quoted above references. The second is optimizing the restoring unitary transformation applied to the extended receiver. Both lead to an increase in $|\lambda|^2$. In addition, $|\lambda|^2$ increases with an increase in the  dimensionality of the extended receiver}. 
\item
Any Hamiltonian with arbitrary (but fixed) coupling constants can govern the spin dynamics provided that such dynamics  preserves the excitation number in the spin chain.
\item
{The dimensionality of the initial sender's state to be perfectly transferred is not principal. Increasing the state dimensionality leads just to increase in the dimensionality of  the restoring unitary transformation applied to  the extended  receiver. }
\item
The   ancilla-state measurement  is  the principal final step  of the algorithm.
 \end{enumerate}
 }
 
 {  
 At this stage we have to note  the growing popularity of measurement-based algorithms  in quantum information theory  \cite{USBGSN}.
The most natural purpose of  using the  measurement is obtaining  the classical information from the quantum system, for instance, 
in Shor's \cite{Shor1} and Grover's \cite{Grover} algorithms.
The  variational algorithms  use measurement to extract  an ``intermediate'' result  of the loss-function value at
fixed values of optimization parameters \cite{CABBEF}.  Then, the calculated value of loss function  is used in the optimization loop based on the  classical optimization algorithm.   
The teleportation algorithm \cite{Popescu,HHH},
  error mitigation algorithm  \cite{BGKMSZ}, HHL-algorithm for solving linear systems \cite{HHL} use  the intermediate-measurement result  for further evaluation of the algorithm. 
The algorithms  constructing the eigenvalues of non-unitary matrices in Ref. \cite{WWLN} also implements measurement.  
The measurement of certain (entangled) states is used in  quantum computation \cite{Wei},  in creating remote entangled states via quantum repeater \cite{ZDB}, in quantum communication  over long distance \cite{ZBD},  
in { the initial  state-preparation} algorithm \cite{ZYY}, 
in matrix-algebra  algorithms  \cite{ZQKW_2024, ZBQKW_arXive2024}.
  The quantum control of dynamics of a two-level system  by non-selective von Neumann measurements of the optimal system observables is studied in \cite{PISR}. This method was further developed to control the  three-level \cite{SZPWSR} and multi-level quantum 
dynamics  \cite{SPHR} and the dynamics of open quantum systems \cite{WPRHT}. Measurement-assisted coherent and non-coherent control of the three-level quantum dynamics is studied in \cite{EP}.
 }

The paper is organized as follows. In Sec.\ref{Section:T} we describe the general algorithm for the {   probabilistic} PST applicable to any pure state but including the transformations over the whole communication line. The detailed description of the algorithm for the  {   probabilistic} PST of the $k$-excitation state is given in Sec.\ref{Section:FixEx}. This algorithm involves only the local restoring   transformation  over the extended receiver.  The algorithm for the  {   probabilistic}  PST of an arbitrary state based on encoding this state into the $k$-excitation state is discussed in Sec.\ref{Section:ArbStTr}.  In Sec.\ref{Section:rest}, we give the detailed description of the $k$-excitation state restoring, {which
uses the restoring unitary transformation   preserving   the excitation number  in the quantum system. Numerical simulations of the $k$-excitation state restoring  (which is the core of the {   probabilistic} PST) with single $\lambda$-parameter using the homogeneous (and {weak end-bonds}) spin chains governed by the $XX$-Hamiltonian are presented in Sec.\ref{Section:num}.}  {   In Sec. \ref{Section:restoring}, we briefly discuss the perturbations of the restoring operator $W_{ER}$ and effect of quantum noise on the probabilistic PST.  Implementation  of the probabilistic PST-algorithm on a superconducting  quantum platform is considered therein.} The basic conclusions are given in Sec. \ref{Section:conclusions}. The detailed  state-restoring  algorithm via the unitary transformation of the extended receiver not preserving the excitation number is presented in the Appendix, Sec.\ref{Section:Appendix}.

\section{Algorithm for perfect state transfer}
\label{Section:T}

\begin{figure*}[!]
\centering
\includegraphics[scale=0.65]{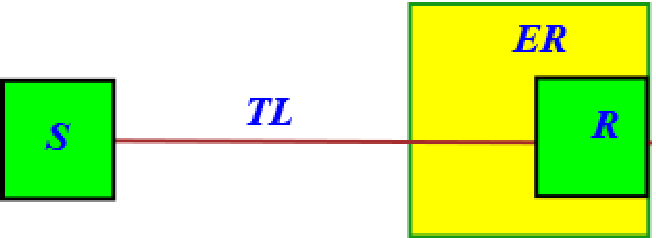}
\caption{Three-partite communication  line including the sender $S$, transmission line $TL$, receiver $R$. The extended receiver $ER$ is also present. {  The pure initial sender's state is transferred to the receiver at some time instant  for state registration $t_0$.}}  
\label{Fig:circuit}
\end{figure*}

{  We study the state transfer along  the three-partite  communication line  including the sender $S$, transmission line $TL$ and receiver $R$, see Fig.\ref{Fig:circuit}. 
The number of spins in   the subsystems  $S$, $TL$ and $R$ is, respectively, $n^{(S)}$, $n^{(TL)}$ and $n^{(R)}$. {Then   $N^{(S)}=2^{n^{(S)}}$,   $N^{(TL)}=2^{n^{(TL)}}$ and $N^{(R)}=2^{n^{(R)}}$ are the numbers of basis states in those subsystems.}
All together, there are $N$ qubits in the communication line, $N= n^{(S)}+n^{(R)}+n^{(TL)}$.   For applying the restoring unitary operator,  we select the extended receiver $ER$ \cite{FPZ_2021,BFLP_2022} which includes  the spins of the receiver   $R$ together with several spins of the transmission line $TL$. These several spins are  called $A$ (ancilla), and the transmission line without the subsystem $A$ is called $TL'$.  
Thus, $ER=A\cup R$, $TL=TL'\cup A$. 
}

In this section, we consider 
the algorithm for PST including  two subroutines. First of them provides the state  transfer from the sender $S$ to  the receiver $R$ via the transmission line $TL$ and state restoring using the unitary transformation applied to the extended receiver $ER$. This subroutine maps  the initial state of the system $S\cup TL\cup R$ (that is the tensor product of an arbitrary pure  sender state and the ground state of $TL\cup R$) into the superposition state including the restored state of the receiver  and garbage. The second subroutine removes the garbage thus leaving only the restored state   which  is  the tensor product of the  ground state of $S\cup TL$ and the receiver state that coincides with the initial sender state.  Thus, these two subroutine provide the perfect transfer of an arbitrary state from the sender to the receiver. {  Now we consider the two above subroutines in more details.}

\subsection{Subroutine 1: state transfer and restoring}
We consider the evolution of a pure quantum state in a communication line constructed on the basis of a spin-1/2 chain whose dynamics is governed by some Hamiltonian $H$ preserving the excitation number in a  quantum state. 
The evolution of a pure state  
 is described by the Schr\"odinger  equation ($\hbar=1$)
\begin{eqnarray}\label{rho00}
i\frac{d}{d t}  |\Psi(t)\rangle&=& H|\Psi(t)\rangle, \;\; [H,I_z]=0,
\end{eqnarray}
where $I_z=\sum_j I_{jz}$ is the $z$-projection of the  total spin momentum {  and $I_{jz}$ is the $z$-projection of the $j$th spin momentum.}  Then
\begin{eqnarray}\label{evol}
 |\Psi(t)\rangle =V(t)  |\Psi(0)\rangle,\;\;V(t)=e^{-iH t}.
\end{eqnarray}
The state preserving condition requires the block-diagonal structure of  the operators $H$ and $V$ in the bases ordered by the excitation number:
\begin{eqnarray}\label{HV}
H={\mbox{diag}}(H^{(0)},H^{(1)},\dots,H^{(N)}),\;\; V={\mbox{diag}}(V^{(0)},V^{(1)},\dots,V^{(N)}).
\end{eqnarray}
{   Thus, the blocks $H^{(i)}$ and $V^{(i)}$ operate in the $i$-excitation subspace, {  i.e., the subspace spanned by the vectors
$|n_1,\dots,n_i\rangle$, where $n_j$, $j=1,\dots,i$, is the position of the $j$th excited spin.} Recall that $0$- and $N$-excitation blocks are scalars, { because each of the associated  subspaces is  spanned by the single vector, respectively, $|0\rangle$ and $|\underbrace{1,\dots,1}_N\rangle$.}
For the problem of state transfer,  the initial state $|\Psi(0)\rangle$ of the communication line including the sender $S$, transmission line $TL$ and receiver $R$ has the following structure:
\begin{eqnarray}\label{PsiIn}
&&
|\Psi(0)\rangle = |\psi(0)\rangle_S |0\rangle_{TL,R},
\end{eqnarray}
where $ |\psi(0)\rangle_S$ is the initial state of the sender,
\begin{eqnarray}
\label{psiS}
&&
 |\psi(0)\rangle_S =\sum_{j=0}^{N^{(S)}-1} s_j |j\rangle_S, \;\;\sum_{j=0}^{N^{(S)}-1} |s_j|^2=1,\;\; N^{(S)}=2^{n^{(S)}},
\end{eqnarray}
and $|0\rangle_{TL,R}$ is the  ground initial state of the transmission line and receiver. 
At some time instant $t_0$ (time instant for state registration) we apply the unitary transformation $W(\varphi)$ to the extended receiver, 
\begin{eqnarray}\label{WW}
W(\varphi)=I_{S,TL'}\otimes W_{ER}(\varphi),
\end{eqnarray} 
\begin{eqnarray}\label{Psi0}
 |\Psi(t_0,\varphi)\rangle =W(\varphi)V(t_0)|\Psi(0)\rangle, 
\end{eqnarray}
where 
$\varphi$ is the list of all parameters in the unitary transformation $W_{ER}$. {   These parameters serve to establish the desired state restoring.  They must be fixed to satisfy the restoring conditions, see 
Sec.\ref{Section:rest}, Eqs. (\ref{TT1}), (\ref{TT2}).}
We require that $W$ preserves the excitation number, i.e., 
\begin{eqnarray}\label{Wcom}
[W,I_z]=0.
\end{eqnarray}
Then the operators $W_{ER}$ and $W$ have the block diagonal structure as well,
\begin{eqnarray}\label{WW20}
&&W_{ER}={\mbox{diag}}(W^{(0)}_{ER},W^{(1)}_{ER},\dots,W^{(n^{(ER)})}_{ER} ),\\\label{WW2}
&&W={\mbox{diag}}(W^{(0)},W^{(1)},\dots,W^{(N)}),
\end{eqnarray}
where only blocks governing up to $n^{(ER)}<N$ excitations (number of qubits in the extended receiver) are nontrivial, all other blocks are identities. 
The structure of  the  pure state $|\Psi(t_0,\varphi)\rangle$ at some fixed time instant $t_0$ is as follows:
\begin{eqnarray}\label{Phi1}
|\Phi_1(t_0,\varphi)\rangle\equiv  |\Psi(t_0,\varphi)\rangle = |0\rangle_{S,TL} |\psi(t_0,\varphi)\rangle_R + |g_1\rangle,
\end{eqnarray}
where the state of the receiver $|\psi(t_0,\varphi)\rangle_R$ reads {  (for still arbitrary parameters $\varphi$)}
\begin{eqnarray}\label{rj}
|\psi(t_0,\varphi)\rangle_R = \sum_{j=0}^{N^{(R)}-1} r_j(t_0,\varphi) |j\rangle_R.
\end{eqnarray} 
In Eq.(\ref{Phi1}), the first part contains the useful terms, while the  non-normalized state $|g_1\rangle$, $\langle g_1|g_1\rangle <1$,  includes all the rest terms and it is  called garbage. Hereafter, the garbage state is orthogonal to the useful part of the superposition state.
Due to the presence of this garbage the probability amplitudes $r_j$ satisfy the inequality 
\begin{eqnarray}
 \sum_{j=0}^{N^{(R)}-1} |r_j|^2<1. 
\end{eqnarray}
Remark, that the first term in Eq.(\ref{Phi1}) is always present in the   whole superposition state. 
 Below we show how this term can be extracted from the general superposition state assuming  $N^{(S)}=N^{(R)}$, i.e., the number of qubits in the sender and receiver is the same, $n^{(R)}=n^{(S)}$.
 
The state restoring \cite{FPZ_2021} means finding such  parameters  $\varphi_0$ of the transformation $W_{ER}(\varphi)$ that  the restored structure of $|\psi(t_0,\varphi_0)\rangle_R$ has  the form
\begin{eqnarray}\label{restore}
 |\psi(t_0,\varphi_0)\rangle_R  = \sum_{j=0}^{N^{(R)}-1}  \lambda_j s_j  |j\rangle_R,
\end{eqnarray}
i.e., each probability amplitude $r_j$ in Eq.(\ref{rj}) is proportional to the appropriate probability amplitude $s_j$ with the parameter $\lambda_j$ independent on the initial sender state (i.e., independent on the parameters $s_k$, $k=0,\dots, N^{(S)}-1$).  
For the purpose of PST, we require that all $\lambda$-parameters equal each other: $\lambda_j =\lambda$, $\forall j$. Thus the state $|\Phi_1\rangle$ in Eq. (\ref{Phi1})  after restoring reads:
\begin{eqnarray}\label{Phi1star}
&&
|\Phi_1\rangle\equiv  |\Psi(t_0,\varphi_0)\rangle=W(\varphi_0)V(t_0)|\Psi(0)\rangle = |0\rangle_{S,TL} |\psi(t_0,\varphi_0)\rangle_R  + |g_1\rangle, \\\label{restore20}
&&
|\psi(t_0,\varphi_0)\rangle_R  = \lambda \sum_{j=0}^{N^{(R)}-1} s_j  |j\rangle_R.
\end{eqnarray}
{ Without loss of generality, we set $\lambda>0$ hereafter, because the phase factor can be removed by the unitary transformation $W_{ER}$.}
 We note that the state  $|\psi(t_0,\varphi_0)\rangle_R$ in Eq. (\ref{restore20}) is not normalized because it is a part of the state $|\Phi_1\rangle$ in Eq.(\ref{Phi1star}). {   The restoring process is described in Sec.\ref{Section:rest} in details.}

It remains to select and remove the garbage. However, this is not a simple problem because we would like  to resolve it using only the local manipulations at the extended  receiver  leaving the nodes of the sender and transmission line untouched. This is done in   Secs.\ref{Section:FixEx} and \ref{Section:ArbStTr}. Here we just show how the problem of garbage removal  can be resolved if we disregard the requirement of using only the local operations at the extended receiver { and allow operating with all nodes of the communication line.}

\subsection{Subroutine 2: garbage removal}
\label{Section:arb}\label{Section:OM0}
The obtained superposition state $|\Phi_1\rangle$  in Eq.(\ref{Phi1star}) includes  $|\psi(t_0,\varphi_0)\rangle_R$ in the first  term.  Now,  our purpose is to remove the garbage $|g_1\rangle$ and thus  transform $|\Phi_1\rangle$  to the tensor product state in the form  
$|0\rangle_{S,TL}|\psi(t_0,\varphi_0)\rangle_R$ with the ground  state $|0\rangle_{S,TL}$ of the sender and transmission line.
For that purpose  we introduce the one-qubit ancilla $B$ in the ground state, the projector
\begin{eqnarray}\label{PSTL}
P_{S,TL}=|0\rangle_{S,TL}\, {_{S,TL}\langle 0|}
\end{eqnarray}
and the  controlled operator
\begin{eqnarray}\label{W1STL}
W^{(1)}_{S,TL,B}=P_{S,TL}\otimes \sigma^{(x)}_{B} + (I_{S,TL}-P_{S,TL})\otimes I_B. 
\end{eqnarray}
{  Hereafter, the subscript at the operator indicates  the subsystem to which this operator is applied. In turn, the subscript at the subsystem indicates  the qubit of this subsystem.}
Applying the operator $W^{(1)}_{S,TL,B}$ to $|\Phi_1\rangle$ we obtain
\begin{eqnarray}\label{Phi20}
|\Phi_2\rangle =  W^{(1)}_{S,TL,B}|\Phi_1\rangle = |0\rangle_{S,TL} |\psi(t_0,\varphi_0)\rangle_R |1\rangle_{B}+ |g_0\rangle |0\rangle_B.
\end{eqnarray}
Thus, we label the garbage by the state $|0\rangle_B$. 

{ Now we perform the measurement of the ancilla qubit $B$ with the purpose of removing the garbage $|g_0\rangle$ from the superposition state $|\Phi_2\rangle$. 
By measurement  of the one-qubit ancilla $B$ we call  the  operator $M^{(2)}_{B}$  that,  being applied to the superposition state 
\begin{eqnarray}\label{oneq}
|\Phi\rangle_{B} =\alpha |1\rangle_{B} + \beta |0\rangle_{B},\;\; |\alpha|^2 +  |\beta|^2=1,
\end{eqnarray}
projects  it to one  of the superposed  basis states, either  $|1\rangle_{B}$ or $|0\rangle_{B_2}$, with the probability, respectively, $|\alpha|^2$ and $|\beta|^2$:
 \begin{eqnarray}
 M^{(2)}_{B} |\Phi\rangle_{B} =\left\{
 \begin{array}{ll}\displaystyle
  |1\rangle_{B}, & {\mbox{probability}} \;\; |\alpha|^2\cr\displaystyle
|0\rangle_{B}, & {\mbox{probability}} \;\; |\beta|^2
  \end{array}
 \right. .
 \end{eqnarray}
Therefore, applying $M^{(2)}_{B}$ to the state $|\Phi_2\rangle$ we  get the desired output $|1\rangle_{B}$
 with the probability $\lambda^2$  {  (success probability)}, thus selecting the following state  of the system $S\cup TL \cup R$:}
\begin{eqnarray}\label{Phi3}
&&
|\Phi_3\rangle =
\frac{1}{\lambda} |0\rangle_{S,TL} |\psi(t_0,\varphi_0)\rangle_R = |0\rangle_{S,TL} |\Psi_{out}\rangle  ,\\\label{PsiOut}
&&
|\Psi_{out}\rangle= \frac{1}{\lambda}|\psi(t_0,\varphi_0)\rangle_R  =\sum_{j=0}^{N^{(R)}-1} s_j |j\rangle_R, 
\end{eqnarray} 
where the parameter $\lambda$ disappears { in the right hand side of Eq.(\ref{PsiOut})} due to the normalization condition for the pure 
state 
$|\Phi_3\rangle$. The derived state $|\Psi_{out}\rangle$  coincides with the initial sender state  $|\psi\rangle_S$.
Due to the above probabilistic result of the ancilla-state measurement, we  are interested in the  largest possible parameter $\lambda$.

{    { We also can appeal to the probability amplification theorem \cite{KShV}. 
If we perform $M$ runs of the algorithm, then the only undesired event is when  all $M$ measurements of the ancilla 
$B$ will be unsuccessful. The probability of such event is $(1-\lambda^2)^M$. Thus, if we want this probability  be $\varepsilon \ll 1$, i.e.,
\begin{eqnarray}
(1-\lambda^2)^M =\varepsilon,
\end{eqnarray}
then  we can estimate the required number of runs $M$ as follows:
\begin{eqnarray}
M=\frac{\log_2\frac{1}{\varepsilon}}{\log_2 \frac{1}{1-\lambda^2}}.
\end{eqnarray}
This formula shows that $M$ logarithmically increases with a decrease in $\varepsilon$. However, if $\lambda^2$ is small, then 
\begin{eqnarray}
M\approx \frac{\ln 2}{ \lambda^2} \log_2\frac{1}{\varepsilon}.
\end{eqnarray}
In this case the amplification is less effective because the number of runs not only logarithmically increases with decrease in $\varepsilon$, but  it is also  inversely proportional to the success probability. 
Therefore, we are interested in the largest possible $\lambda$.}}

{   The fact that the algorithm requires $M$  running makes certain constraints on applicability of the algorithm. To effectively apply this algorithm, we must be able to provide $M$ copies of the state to be transferred.  Of course this can not be done for an unknown state due to the no-cloning theorem \cite{KShV}. However, we can deal with $M$-channel computation, each channel includes the probabilistic PST-circuit  as a subroutine with its own ancilla, so that the postselection after all ancillae measurements yields the perfectly transferred desired state.}

{We have to emphasize that,  after getting the desired result of the ancilla  measurement, we obtain the state of receiver $|\Psi_{out}\rangle$,  given in Eq.(\ref{PsiOut}),  coinciding with the initial sender state. In other words, the fidelity of this state transfer is 
exactly one justifying that we deal with  the PST, while the success probability $\lambda^2<1$   defines the probability of getting PST in result of single run of the algorithm. {Repeating the algorithm we increase the probability of  registering  the PST}. This is a principal difference with the high-fidelity state transfer, when the transferred state approximates the initial sender state but does not exactly coincide with it. In that case repeating the algorithm does not increase fidelity.} 

The circuit for the above algorithm of PST is presented in Fig.\ref{Fig:Arb}.
\begin{figure*}[!]
\centering
\includegraphics[scale=0.65]{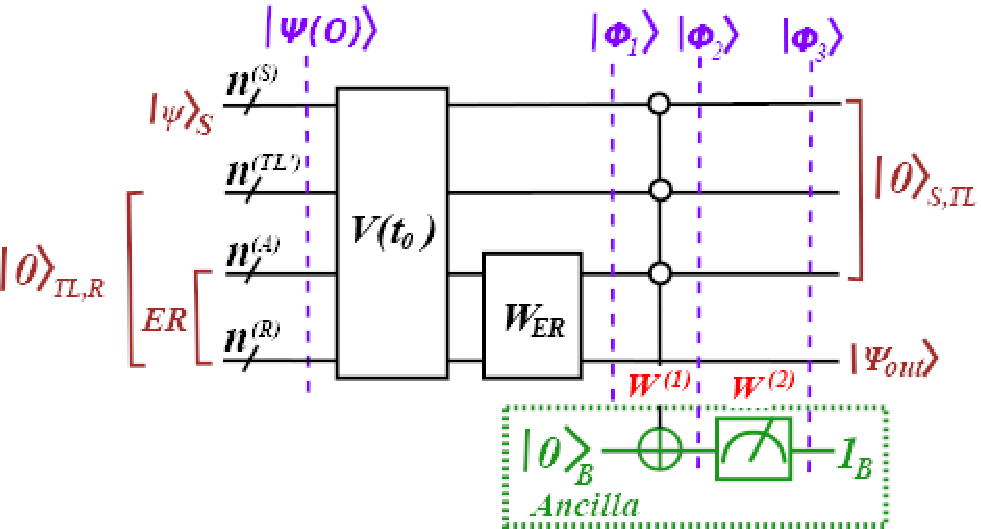} 
\caption{The circuit for the perfect transfer of  an arbitrary pure state. The controlling subsystem  in the operator  $W^{(1)} \equiv W^{(1)}_{S,TL,B}$ includes the qubits of sender $S$ and transmission line $TL$. { Here $|\Psi(0)\rangle$ is the initial state given in Eq.(\ref{PsiIn}),  $|\Phi_j\rangle$, $j=1,2,3$, are the intermediate states obtained in Eqs.(\ref{Phi1star}),  (\ref{Phi20}), (\ref{Phi3}),  $|\Psi_{out}\rangle$ is defined in Eq.(\ref{PsiOut}),  $V$ is the evolution operator given in Eq. (\ref{evol}), $W_{ER}$ is the restoring unitary transformation applied to the extended receiver $ER$,  $W^{(2)}\equiv W^{(2)}_B$ is the measurement.}}  
\label{Fig:Arb}
\end{figure*}

\section{Perfect transfer of pure fixed-excitation state}
\label{Section:FixEx}
The disadvantage of the algorithm  considered in Sec.\ref{Section:T} is the global non-locality of the garbage removal subroutine, because it includes not only the nodes of the extended receiver $ER$ (along with the ancilla $B$) but also the nodes of the sender $S$ and $TL'$, see Sec.\ref{Section:OM0}.  
Now we consider a variant of the PST where the global non-locality of the  garbage removal is overcome. This is possible for the perfect transfer  of a pure   $k$-excitation state (i.e., the superposition state in which each term has exactly $k$ excited qubits), where both the state restoring and garbage removal subroutines are based  on just the local transformations of the extended receiver. 

\subsection{Subroutine 1: state transfer and restoring}

We enumerate all $k$-excitation states as
\begin{eqnarray}\label{setS}
|\chi_j\rangle_S, \;\;j=0,\dots N^{(S)}_k-1, \;\;N^{(S)}_k = \binom{n^{(S)}}k.
\end{eqnarray}
Thus, in the initial state (\ref{PsiIn}), we have
\begin{eqnarray}\label{Psik}
|\psi(0)\rangle_S = \sum_{j=0}^{N^{(S)}_k -1} s_j |\chi_j\rangle_S, \;\;\sum_{j=0}^{N^{(S)}_k -1} |s_j|^2=1.
\end{eqnarray}
After transferring the initial sender state  along the communication line,
the final state of the whole system $|\Phi_1\rangle$, defined in Eq.(\ref{Phi1}), becomes 
\begin{eqnarray}\label{Phi1star2}
&&
|\Phi_1(t_0,\varphi)\rangle\equiv  |\Psi(t_0,\varphi)\rangle =W(\varphi)V(t_0)|\Psi(0)\rangle = |0\rangle_{S,TL}|\psi(t_0,\varphi)\rangle _R + |g_1\rangle.
\end{eqnarray}
Substituting the restoring values $\varphi_0$ for the  parameters $\varphi$ we obtain
\begin{eqnarray}
\label{restore2}
&&
|\psi(t_0,\varphi_0)\rangle _R=\lambda  \sum_{j=0}^{N^{(S)}_k-1} s_{j}  |\chi_j\rangle_{R}.
\end{eqnarray}

\subsection{Subroutine 2: garbage removal}\label{Section:OM}
Now, to label and remove the garbage  in $|\Phi_1\rangle$, given in Eq. (\ref{Phi1star2}) with $\varphi=\varphi_0$,  {   we proceed as follows. 
At first, we introduce the   projectors
\begin{eqnarray}\label{PRj}
P_{\tilde R^{(j)}} =|\chi_j\rangle_{\tilde R^{(j)}} \;{_{\tilde R^{(j)}}\langle\chi_j|} , \;\; j=0, \dots, N^{(S)}_k-1,
\end{eqnarray}
where  $\tilde R^{(j)}$ is the state-subspace including the states spaces of the excited qubits in the state  $ |\chi_j\rangle_{R}$.     Then $ |\chi_j\rangle_{\tilde R^{(j)}} $  is the tensor product of the $k$ excited  states corresponding to the excited qubits in 
$|\chi_j\rangle_R$.  For instance, if $k=2$ and $|\chi_2\rangle_R=|101\rangle_R$, then 
$|\chi_2\rangle_{\tilde R^{(2)}}=|1\rangle_{R_1} |1\rangle_{R_3}$. 
In addition, we introduce the 1-qubit ancilla $B$ in the ground state and  construct the controlled  operator based on the projectors (\ref{PRj}):
\begin{eqnarray}\label{W1}
\label{W1RB1}\label{W1B}
&&
W^{(1)}_{RB}=\prod_{j=0}^{N^{(S)}_k-1} \Big(P_{\tilde R^{(j)}}   \otimes \sigma^{(x)}_{B} +  (I_{\tilde R^{(j)}}-P_{\tilde R^{(j)}}) \otimes I_{B}\Big).
\end{eqnarray}}
Applying the operator $W^{(1)}_{RB}$  to $|\Phi_1(t_0,\varphi_0)\rangle |0\rangle_B$ we obtain
\begin{eqnarray}\label{Phi2}
|\Phi_2\rangle =  W^{(1)}_{RB}|\Phi_1\rangle |0\rangle_B = |0\rangle_{S,TL} |\psi(t_0,\varphi_0)\rangle_R |1\rangle_{B}+ |g_2\rangle |0\rangle_B.
\end{eqnarray}
{ The depth of the operator $W^{(1)}_{RB}$ is defined {    by the number of  the  projectors $P_{\tilde R^{(j)}}$  and their depth}  and equals 
\begin{eqnarray}\label{dW1}
O(k N^{(S)}_k),
\end{eqnarray}
 where we take into account that  the state $|\chi_j\rangle_{R_j}$ includes $k$ excited one-qubit states and 
  assume that the depth of the controlled operator with $k$ controlling qubits is $O(k)$ \cite{KShV}. 
  Now, to remove the garbage, we measure the state of the ancilla $B$ to get the desired  output $|1\rangle_{B}$ with the probability $\lambda^2$. }
Then we arrive at the state 
\begin{eqnarray}\label{Phi32}
&&
|\Phi_3\rangle = 
 \frac{1}{\lambda} |0\rangle_{S,TL} |\psi(t_0,\varphi_0)\rangle_R =
 |0\rangle_{S,TL} |\Psi_{out}\rangle,\\\label{PsiOut0}
 &&
  |\Psi_{out}\rangle=\sum_{j=0}^{N^{(S)}_k-1} s_{j}|\chi_j\rangle_{R}
 ,
\end{eqnarray} 
where $ |\psi(t_0,\varphi_0)\rangle$ is given in Eq.(\ref{restore2}) 
and  the parameter $\lambda$ disappears from $|\Psi_{out}\rangle$ due to the normalization condition for the pure 
state 
$|\Phi_3\rangle$, so that the output state $|\Psi_{out}\rangle$ coincides with $|\psi\rangle_S$ in Eq.(\ref{Psik}). Of course, the probability amplification considered in Sec.\ref{Section:OM0} is applicable in this section as well.

The time $T^{(PST)}$ required for performing  the   PST-algorithm is defined by the time instance for state registration $t_0$ (this is the time interval needed for the state transfer along the spin chain), by the depth of the operator $W_{ER}$, which will be  discussed in Sec.\ref{Section:rest}, and by the  depth  of the operator $W^{(1)}_{RB}$ given  in Eq.(\ref{dW1}).  In addition, due to the probabilistic removing  the garbage via the ancilla measurement, the depth will be related to  the number of runs $N^{(run)}$  of the algorithm needed to get access to the excited state of the ancilla $B$, $N^{(run)}=O(\lambda^{-2})$. 
Thus, introducing the time interval $t^{(op)}$ for single quantum operation, we estimate   $T^{(PST)}$ as follows: 
\begin{eqnarray}
T^{(PST)} =O \left(\Big(t_0 + t^{(op)} \big({\mbox{depth}}(W_{ER}) + k N^{(S)}_k\big)\Big) N^{(run)}\right).
\end{eqnarray}
The circuit for the algorithm of $k$-excitation PST  is shown in Fig.\ref{Fig:1ex}.
\begin{figure*}[!]
\centering
\includegraphics[scale=0.55]{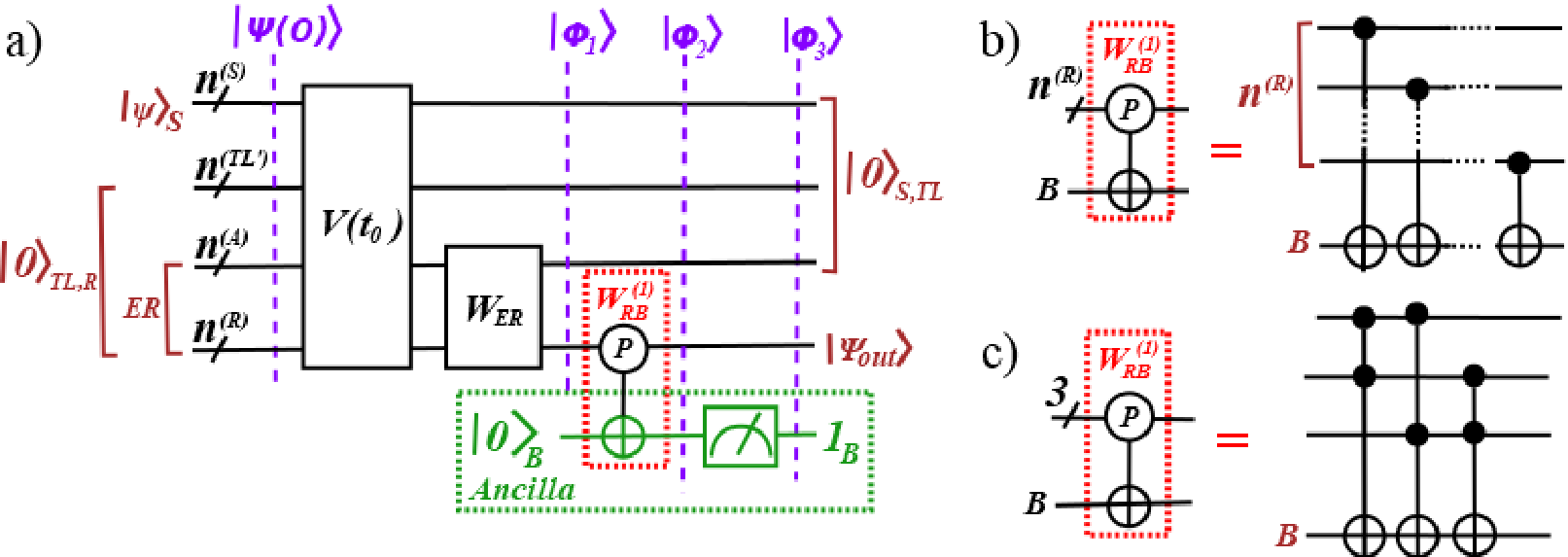} 
\caption{(a) The circuit for the perfect transfer of  an arbitrary   $k$-excitation pure state. {  Here $|\Psi(0)\rangle$ is the initial state (\ref{PsiIn}) with the sender's state (\ref{Psik}), $|\Phi_j\rangle$, $j=1,2,3$, are the intermediate states given in Eqs.(\ref{Phi1star2}), (\ref{Phi2}) and  (\ref{Phi32}), the output state $|\Psi_{out}\rangle$ is given in Eq.(\ref{PsiOut0}).} Two examples of circuits  for the controlled operator $W^{(1)}_{RB}$  given in Eq.(\ref{W1RB1}) are shown  (b) {  for the 1-excitation state transfer ($k=1$) and the receiver with arbitrary $n^{(R)}$  and (c) for the 2-excitation state transfer ($k=2$) and the three-qubit receiver  ($n^{(R)}=3$).} }  
\label{Fig:1ex}
\end{figure*}
Examples of the circuit  for $W^{(1)}_{RB}$ in the case of  1- and 2-excitation states are  shown, respectively,  in Fig.\ref{Fig:1ex}b and Fig.\ref{Fig:1ex}c. In the later case, we take the 3-qubit receiver.

\section{Perfect transfer of arbitrary pure state}
\label{Section:ArbStTr}
We demonstrated in Sec.\ref{Section:FixEx}   that the $k$-excitation state can be perfectly transferred using evolution and local manipulations on the extended receiver. 
This fact proposes the following strategy for the perfect transfer of an arbitrary state. First, we encode an arbitrary state of some subsystem $S^{(0)}$, $|\psi(0)\rangle_{S^{(0)}}$, into the $k$-excitation state of the sender $S$. Then transfer this state to the receiver $R$ and restore the transferred state. Finally, we  decode the $k$-excitation state to the state of some subsystem $R^{(0)}$. After  removing the garbage via the ancilla measurement, this state  coincides with $|\psi(0)\rangle_{S^{(0)}}$.

Consequently, the PST-algorithm consists of five subroutines.
\begin{enumerate}
\item 
{\it Subroutine 1:}  Map  the initial arbitrary $n^{(S^{(0)})}$-qubit  state $|\psi(0)\rangle_{S^{(0)}}$ of the subsystem $S^{(0)}$  into  the $k$-excitation state of the sender $S$  having  slightly larger dimension.
 \item
{\it Subroutine 2:}   Subject this $k$-excitation  state to the evolution and state restoring at the receiver fixing the time instant for state registration  $t_0$ and parameters $\varphi_0$ of the restoring unitary transformation $W_{ER}$. Now the superposition state includes the restored  receiver's state that coincides with the initial $k$-excitation state of the sender $S$ up to the scale factor $\lambda$. The rest of the superposition state is called  garbage.
\item
{\it Subroutine 3:}  Introduce the ancilla to select  the  useful information   (the $k$-excitation restored state of the receiver) from the superposition state. 
 \item
{\it Subroutine 4:}  Map the restored and labeled $k$-excitation state   of the receiver $R$  to the $n^{(R^{(0)})}$-qubit  state of the subsystem $R^{(0)}$, $|\psi(t_0,\varphi_0)\rangle_{R^{(0)}}$, $n^{(R^{(0)})}= n^{(S^{(0)})}$, coinciding with the initial state of $S^{(0)}$, $|\psi(0)\rangle_{S^{(0)}}$, up to the scale factor $\lambda$.  
\item
{\it Subroutine 5:}
Remove the garbage  via the ancilla measurement {thus obtaining the state of $R^{(0)}$ coinciding with the initial state of $S^{(0)}$, $|\psi(0)\rangle_{S^{(0)}}$}. This operation completes the PST-algorithm.
\end{enumerate}
To evaluate this algorithm we have to modify the 
communication line as shown in  Fig.\ref{Fig:circuit2}. Again, it includes
the sender $S$, receiver $R$ and transmission line $TL$ connecting them. But now we also select the subsystems $S^{(0)}$ and $R^{(0)}$ consisting of  {certain numbers of}  nodes of the $TL$ {neighboring to, respectively, $S$ and $R$}. These two subsystems serve to encode the state to be transferred (subsystem $S^{(0)}$) and to register the  transferred state (subsystem $R^{(0)}$). The extended receiver $ER$ includes the nodes of the receiver $R$ and several nodes of the transmission line. At that, the subsystem $R^{(0)}$ is included (at least partially) in  $ER$. We also call $TL''$ the part of $TL$ without nodes of the subsystems $S^{(0)}$ and $R^{(0)}$.

\begin{figure*}[!]
\centering
\includegraphics[scale=0.65]{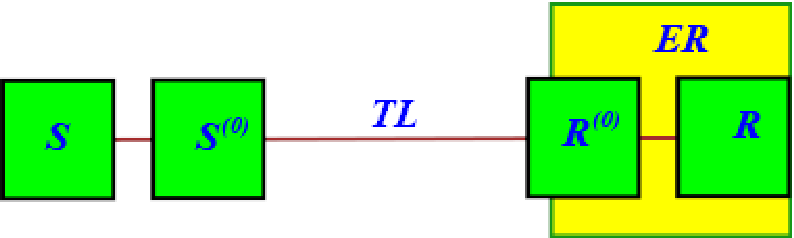} 
\caption{Communication  line  for the perfect transfer of an  arbitrary  pure state. The state to be transferred  and transferred states are encoded into, respectively, the subsystems $S^{(0)}$ and $R^{(0)}$.  {  The arbitrary state of $S^{(0)}$ is encoded into the $k$-excitation state of $S$, this state is perfectly  transferred along the transmission line $TL$ to the receiver $R$ at some time instant for state registration  $t_0$, then the transferred $k$-excitation state  is decoded to the state of $R^{(0)}$ coinciding with the  initial state of $S^{(0)}$. The extended receiver $ER$ includes all nodes of the receiver $R$ and at least several nodes of the subsystem $R^{(0)}$.}}  
\label{Fig:circuit2}
\end{figure*}

Thus, since the $k$-excitation state can be perfectly transferred, we encode an arbitrary $n^{(S^{(0)})}$-qubit state  of the subsystem $S^{(0)}$ into the $k$-excitation state of the larger subsystem $S$. 
We can  estimate, how larger  the subsystem $S$ must be in comparison with the subsystem $S^{(0)}$. 

{Let  us encode an arbitrary state of the $n$-qubit system into the $k$-excitation state of some $m$-qubit system.}  The number of states in the $n$-qubit system equals 
\begin{eqnarray}
\sum_{k=0}^n\binom{n}k = 2^n.
\end{eqnarray}
For the given $m$, the maximal number of $k$-excitation states in the $m$-qubit system corresponds to $k=\lceil \frac{m}{2} \rceil$.
For simplicity, we consider even  $m$.  Then the following asymptotic formula holds:
\begin{eqnarray}
 \binom{m}{m/2} \to  \frac{2^{m}}{\sqrt{\pi m/2}}>\frac{2^m}{
 \sqrt{2 m}}=2^{m-\frac{1}{2}(\log_2m+1)} \;\;{\mbox{as}}\;\; m\to \infty.
\end{eqnarray}
To encode  an arbitrary  $n$-qubit pure state into the $m/2$-excitation state of the $m$-qubit system we need {(for large $n$ and $m$)}
\begin{eqnarray}
2^n \le \binom{m}{m/2}\;\;\Rightarrow\;\;
n<m-\frac{1}{2}\log_2 m-\frac{1}{2}.
\end{eqnarray}
Therefore, $m$ is not significantly larger than $n$ for large systems, {i.e., the number of qubits in  $S$ does not significantly exceed the number of qubits in $S^{(0)}$.} {The same holds for the subsystems $R$ and $R^{(0)}$.}

{Now we give the detailed description of the above five subroutines.}

\subsection{Subroutine 1: encoding arbitrary state into $k$-excitation state}
\label{Section:encoding}
Let us consider an arbitrary $n^{(S^{(0)})}$-qubit pure state of the subsystem $S^{(0)}$:
\begin{eqnarray}
|\Psi\rangle_{S^{(0)}}= \sum_{j=0}^{N^{(S^{(0)})} -1} s_j |j\rangle_{S^{(0)}}
\end{eqnarray}
and the $n^{(S)}$-qubit subsystem $S$ in the ground state. Thus the state of the whole communication line  reads
\begin{eqnarray}\label{Phi00}
|\tilde\Phi_0\rangle = |0\rangle_{S} |\Psi\rangle_{S^{(0)}} |0\rangle_{TL'',R^{(0)},R}.
\end{eqnarray}
We enumerate the $k$-excitation states of $S$ according to Eq.(\ref{setS}).
We also enumerate the basis elements of $S^{(0)}$ 
\begin{eqnarray}
|j\rangle_{S^{(0)}},\;\; j =0,\dots,N^{(S^{(0)})} -1.
\end{eqnarray}
Assuming that 
\begin{eqnarray}\label{NN}
 N^{(S^{(0)})} \le N^{(S)}_k  ,
\end{eqnarray}
 we consider the encoding as  the map
\begin{eqnarray}\label{map}
|j\rangle_{S^{(0)}}  \to |\chi_j\rangle_{S},\;\; j=0,\dots,N^{(S^{(0)})} -1.
\end{eqnarray}
{ We note that not all states from set (\ref{setS}) are used in map (\ref{map}) in general due to the inequality 
(\ref{NN}).}
Map  (\ref{map}) can be realized by the set of  controlled operators 
\begin{eqnarray}\label{tildeWj}
\overline W_{S^{(0)}S}^{(j)}=P^{(j)}_{S^{(0)}} \otimes \sigma^{(x)}_{\chi_j} + (I_{S^{(0)}}-P^{(j)}_{S^{(0)}})\otimes I_{S},\;\; j=0,\dots, N^{(S^{(0)})} -1,
\end{eqnarray}
based on the projectors
\begin{eqnarray}
P^{(j)}_{S^{(0)}}=|j\rangle_{S^{(0)}} \, {_{S^{(0)}}\langle j|},  \;\;j=0,\dots,N^{(S^{(0)})} -1.
\end{eqnarray}
 In Eq.(\ref{tildeWj}),  $\sigma^{(x)}_{\chi_j}$ is the set of $k$ operators $\sigma^{(x)}$ that must be  applied to the qubits of the  subsystem $S$ in the  ground state to create the state $|\chi_j\rangle_{S}$:  $\sigma^{(x)}_{\chi_j}|0\rangle_{S}= |\chi_j\rangle_{S}$. {  
 Since  the operator $\overline W_{S^{(0)}S}^{(j)}$  includes $k$ operators $\sigma^{(x)}$ and $n^{(S^{(0)})}$ controlling qubits (in the projector $P^{(j)}_{S^{(0)}}$),   the depth of this operator is $O(k n^{(S^{(0)})})$. }

We collect the operators $\overline W_{S^{(0)}S}^{(j)}$ in
\begin{eqnarray}
\tilde W^{(0)}_{S^{(0)}S} = \prod_{j=0}^{N^{(S^{(0)})} -1} \overline W_{S^{(0)}S}^{(j)},
\end{eqnarray}
{ whose depth is 
\begin{eqnarray}\label{d1}
O(N^{(S^{(0)})} k n^{(S^{(0)})}).
\end{eqnarray}}
Thus
\begin{eqnarray}\label{tPhi10}
|\tilde \Phi_1\rangle =\tilde  W^{(0)}_{S^{(0)}S} |\tilde \Phi_0\rangle =
 \sum_{j=0}^{N^{(S^{(0)})} -1} s_j  |\chi_j\rangle_{S} |j\rangle_{S^{(0)}}|0\rangle_{TL'',R^{(0)},R},
\end{eqnarray}
{   Now, we put the  subsystem $S^{(0)}$ to the ground state. For this purpose we introduce another set of the
controlled operators 
\begin{eqnarray}\label{tildeWj2}
\overline W_{SS^{(0)}}^{(j)}=P_{\tilde S^{(j)}} \otimes \sigma^{(x)}_{j} + (I_{\tilde S^{(j)}}-P_{\tilde S^{(j)}})\otimes I_{S^{(0)}},\;\; j=0,\dots, N^{(S^{(0)})} -1,
\end{eqnarray}
with the projectors  $P_{\tilde S^{(j)}} $,
\begin{eqnarray}\label{PjS}
P_{\tilde S^{(j)}} = |\chi_j\rangle_{\tilde S^{(j)}}\;{_{\tilde S^{(j)}}\langle\chi_j|}, \;\;  j=0,\dots, N^{(S^{(0)})}-1,
\end{eqnarray}
where the state-subspace $\tilde S^{(j)}$  includes the state-spaces of the excited qubits in the state $|\chi_j\rangle_S$, similar to the  subspace $\tilde R^{(j)}$ in   Eq.(\ref{PRj}).
Then   $|\chi_j\rangle_{\tilde S^{(j)}}$  is the tensor product of the $k$ excited states corresponding to the excited qubits in 
$|\chi_j\rangle_S$.}
The operator  $\sigma^{(x)}_{j}$ in (\ref{tildeWj2})  is the set of $\sigma^{(x)}$-operators  that must be  applied to the qubits of $S^{(0)}$ in the ground state  to form the state $|j\rangle_{S^{(0)}}$. Then $\sigma^{(x)}_{j}|j\rangle_{S^{(0)}} = |0\rangle_{S^{(0)}}$. Since, in general,  $\sigma^{(x)}_{j}$  is applied to $n^{(S^{(0)})}$ qubits and $\overline W_{SS^{(0)}}^{(j)}$ has $k$ controlling qubits, the depth of $\overline W_{SS^{(0)}}^{(j)}$ is $O(k n^{(S^{(0)})})$.   
We collect the operators $\overline W_{SS^{(0)}}^{(j)}$ in
\begin{eqnarray}
\tilde W^{(1)}_{SS^{(0)}} = \prod_{j=0}^{N^{(S^{(0)})} -1} \overline W_{SS^{(0)}}^{(j)},
\end{eqnarray}
 whose depth  equals to the depth of $\tilde W^{(0)}_{S^{(0)}S}$ given in Eq.(\ref{d1}). Thus,
 \begin{eqnarray}\label{Psi000}
&&
|\tilde \Phi_2\rangle =\tilde W^{(1)}_{SS^{(0)}}|\tilde \Phi_1\rangle  =  \sum_{j=0}^{N^{(S^{(0)})} -1} s_j |\chi_j\rangle_{S} |0\rangle_{S^{(0)}}   |0\rangle_{TL'',R^{(0)},R}=\\\nonumber
&&
\sum_{j=0}^{N^{(S^{(0)})} -1} s_j |\chi_j\rangle_{S} |0\rangle_{TL,R}  \equiv|\Psi(0)\rangle  .
\end{eqnarray}
{ The depth of the circuit  for encoding algorithm is determined by the depths of 
$\tilde W^{(0)}_{S^{(0)}S}$ and $\tilde W^{(1)}_{S^{(0)}S}$ and equals $O(N^{(S^{(0)})} k  n^{(S^{(0)})})$.
Thus, we obtain the initial state 
 $|\Psi(0)\rangle$ that appears in the algorithm in Sec.\ref{Section:T} and is defined in Eqs.(\ref{PsiIn}), (\ref{Psik}).
{ We note that the number of terms { in  the sum of} Eq.(\ref{Psi000}) is $N^{(S^{(0)})}$ rather than $N^{(S)}_k$.}

\subsection{Subroutine 2: state transfer and restoring}
\label{Section:evrest}
Now we run the evolution and successive state restoring applying the operator $W(\varphi) e^{-i Ht_0}$ to the  state $|\Psi(0)\rangle$, defined in Eq. (\ref{Psi000}), and obtaining the state $|\Phi_1(t_0,\varphi)\rangle$, given in  Eq.(\ref{Phi1star2}). The state restoring yields the proper parameters $\varphi=\varphi_0$.
{The time required for evaluation of the  operator  $V(t_0)$ is $t_0$, while 
the depth of $W_{ER}$ included into the operator $W$, given in  Eq.(\ref{WW}), will be discussed in Sec.\ref{Section:rest}.}

\subsection{Subroutine 3: selecting useful information}
\label{Section:select}
{   To select useful information and remove the garbage,}  we introduce the one-qubit ancilla $B$ in the ground state and construct the operator $W^{(1)}_{RB}$, defined in Eq.(\ref{W1RB1}) {   with replacing the upper limit $N^{(S)}_k -1 \to N^{(S^{(0)})}-1$, i.e., 
\begin{eqnarray}\label{W1k}
&&
W^{(1)}_{RB}=\prod_{j=0}^{N^{(S^{(0)})}-1} \Big(P_{\tilde R^{(j)}}   \otimes \sigma^{(x)}_{B} +  (I_{\tilde R^{(j)}}-P_{\tilde R^{(j)}}) \otimes I_{B}\Big).
\end{eqnarray}
 where  the projectors $P_{\tilde R^{(j)}}$ are  defined as
\begin{eqnarray}\label{PRj2}\label{PRj22}
P_{\tilde R^{(j)}} =|\chi_j\rangle_{\tilde R^{(j)}} \;{_{\tilde R^{(j)}}\langle\chi_j|} , \;\; j=0, \dots, N^{(S^{(0)})}-1,
\end{eqnarray}
similar to  the projectors in Eq.(\ref{PRj}). Applying $W^{(1)}_{RB}$} to the state  $|\Phi_1(t_0,\varphi_0)\rangle|0\rangle_{B}$ 
we  label the  first term in Eq.(\ref{Phi1star2}):
\begin{eqnarray}\label{Phi200}
|\Phi_2\rangle = W^{(1)}_{RB}|\Phi_1(t_0,\varphi_0)\rangle|0\rangle_{B}= |0\rangle_{S,TL} |\psi(t_0,\varphi_0\rangle_R |1\rangle_{B} + |g_2\rangle |0\rangle_{B} .
\end{eqnarray}
The depth of  $W^{(1)}_{RB}$ is
\begin{eqnarray}\label{dW11}
O(k N^{(S^{(0)})}).
\end{eqnarray}
 {  Unlike the algorithm in Sec.\ref{Section:FixEx}, we do not apply any measurement in this subroutine.}
 
 {    We note that the  depth of the operator labeling the garbage indicated in Eqs.(\ref{dW1}) and (\ref{dW11}) exponentially increases with the dimension of the sender. This   is the price for the {\it  perfect} (although probabilistic) state transfer of an arbitrary {\it multiqubit} $k$-excitation state.  The advantage of the probabilistic PST is that all adjustment of the algorithm are concentrated at the  receiver side and do not include the nodes of the transmission line $TL'$ unlike the deterministic PST where all coupling constants must be exactly adjusted. Moreover, the usual deterministic PST-algorithms work well for the one-qubit state transfer and require phase-adjustments  of the probability amplitudes of the superposition state at the receiver side for the  multiqubit PST. }

\subsection{Subroutine 4: decoding $k$-excitation state}
\label{Section:decoding}
Let us decode the $k$-excitation state of $R$  into the state of $n^{(R^{(0)})}$-qubit subsystem $R^{(0)}$, $n^{(R^{(0)})}=n^{(S^{(0)})}$,  thus performing the map
\begin{eqnarray}
|\chi_j\rangle_{R}  \to |j\rangle_{R^{(0)}},\;\; j=0,\dots,N^{(S^{(0)})}-1.
\end{eqnarray}
This map can be realized by the set of  controlled operators (similar to operators (\ref{tildeWj2}))
\begin{eqnarray}\label{WRR0}
\overline W_{RR^{(0)}}^{(j)}=P_{\tilde R^{(j)}} \otimes \sigma^{(x)}_{j} + (I_{\tilde R^{(j)}}-P_{\tilde R^{(j)}})\otimes I_{R^{(0)}},   {   \;\; j=0, \dots, N^{(S^{(0)})}-1,}
\end{eqnarray}
 {   based on the projectors $P_{\tilde R^{(j)}}$ defined in Eq.(\ref{PRj2}).}
In Eq.(\ref{WRR0}),   $\sigma^{(x)}_{j}$ is the same as in  Eq.(\ref{tildeWj2}):  $\sigma^{(x)}_{j}|0\rangle_{R^{(0)}} = |j\rangle_{R^{(0)}}$.
Notice that  the operator $\overline W_{RR^{(0)}}^{(j)}$ is applied to the subsystems $R$ and $R^{(0)}$ without specifying the state of $B$.  Later, in Subroutine 5, we select only terms with the state $|1\rangle_{B}$.
The depth of $\overline W_{RR^{(0)}}^{(j)}$ is $O(k n^{(S^{(0)})})$.
We collect the operators  $\overline W_{RR^{(0)}}^{(j)}$ in the operator $ \displaystyle W^{(2)}_{RR^{(0)}} = \prod_{j=0}^{N^{(S^{(0)})}-1} \overline W_{RR^{(0)}}^{(j)}$, {whose depth is defined in Eq.(\ref{d1}).
Applying $W^{(2)}_{RR^{(0)}}$ to $|\Phi_2\rangle$, we obtain
\begin{eqnarray}\label{Phi33}
|\Phi_3\rangle =  W_{RR^{(0)}}^{(2)} |\Phi_2\rangle=\lambda
 \sum_{j=0}^{N^{(S^{(0)})}-1}  s_j  |0\rangle_{S,S^{(0)},TL''}  |j\rangle_{R^{(0)}} |\chi_j\rangle_{R} |1\rangle_{B}   +  |g_3\rangle |0\rangle_{B}  .
\end{eqnarray}
{Next, we need to put the subsystem $R$ into the ground state. For this purpose we introduce the controlled operators (similar to operators (\ref{tildeWj}))
\begin{eqnarray}\label{WR0R}
\overline W_{R^{(0)}R}^{(j)}=P^{(j)}_{R^{(0)}} \otimes \sigma^{(x)}_{\chi_j} + (I_{R}-P^{(j)}_{R^{(0)}})\otimes I_{R},
\end{eqnarray}
 based on the projectors
\begin{eqnarray}\label{PRj3}
P^{(j)}_{R^{(0)}}=|j\rangle_{R^{(0)}} \, {_{R^{(0)}}\langle j|}   
, \;\;j=0,\dots,N^{(S^{(0)})}-1,
\end{eqnarray}
where  $\sigma^{(x)}_{\chi_j}$ is the same as in Eq.(\ref{tildeWj}): $\sigma^{(x)}_{\chi_j}|0\rangle_{R} = |\chi_j\rangle_R$, then
$\sigma^{(x)}_{\chi_j}|\chi_j\rangle_{R} = |0\rangle_R$.  We collect the operators  $\overline W_{R^{(0)}R}^{(j)}$ in the operator $ \displaystyle W^{(3)}_{R^{(0)}R} = \prod_{j=0}^{N^{(S^{(0)})}-1} \overline W_{R^{(0)}R}^{(j)}$, {$N^{(S^{(0)})} = N^{(R^{(0)})}$},
whose depth is defined in Eq.(\ref{d1}).  Applying $W^{(3)}_{R^{(0)}R}$ to $|\Phi_3\rangle$, we have
\begin{eqnarray}\label{Phi42}
|\Phi_4\rangle =  W_{R^{(0)}R}^{(3)} |\Phi_3\rangle=\lambda
 \sum_{j=0}^{N^{(S^{(0)})}-1}  s_j  |0\rangle_{S,S^{(0)},TL''}  |j\rangle_{R^{(0)}} |0\rangle_{R} |1\rangle_{B} +  |g_4\rangle |0\rangle_{B}  .
\end{eqnarray}
}

{    {\bf Remark.} The depth of the encoding/decoding subroutines  determined in Eq.(\ref{d1}) exponentially increases with the dimension of the sender. These subroutines   are included  into the algorithm for the {\it arbitrary} pure state transfer. However, the $k$-excitation states can be used for organizing the certain type of quantum algorithms as well. For instance, the algorithms of matrix manipulations based on the matrix-encoding into the quantum superposition state \cite{ZQKW_2024}, in principle,  can be adjusted for the $k$-excitation state-subspace. In this case, the $k$-excitation state transfer can be utilized without state-encoding/decoding subroutines.
 Then, one can realize the parallel calculations in different fixed-excitation state-subspaces. 
 In this way, we can avoid implementing the state-encoding/decoding subroutines as well as   utilize  all fixed-excitation state-subspaces of the sender/receiver. }

\subsection{Subroutine 5: garbage removal}
Finally, we measure the state of the ancilla $B$ with the desired output $|1\rangle_{B}$. The success probability  of such measurement is
$\lambda^2$.  
As the result we obtain
\begin{eqnarray}\label{Phi62}
&&
|\Phi_5\rangle = 
|0\rangle_{S,S^{(0)},TL'} |\Psi_{out}\rangle  |0\rangle_{R}   ,\\\label{Psiout0}
&&
|\Psi_{out}\rangle=\sum_{j=0}^{N^{(S^{(0)})}-1} s_j  |j\rangle_{R^{(0)}},
\end{eqnarray}
{i.e., the PST.}
The above success  probability estimates the number of runs  of the whole algorithm  (Subroutines 1 -- 5) needed to access the desired state of $B$, this number is $O(\lambda^{-2})$. 

{ The evaluation time of the total PST-algorithm is defined by the depth of state encoding subroutine  (Sec.\ref{Section:encoding}, Eq.(\ref{d1})), by the time evolution $t_0$ and the depth of $W_{ER}$ (Sec.\ref{Section:evrest}), {  by the depth  of the operator $W^{(1)}_{RB}$, Eq.(\ref{W1RB1}), that labels the garbage   (Sec.\ref{Section:select}, Eq.(\ref{dW11})) }
 and, finally, by the depth of the state decoding subroutine (Sec.\ref{Section:decoding}, Eq.(\ref{d1})).
  In view of multiple running, the time of PST, $T^{(PST)}$, is
\begin{eqnarray}\label{dQ}
&&
T^{(PST)} =\\\nonumber
&& O\left(\Big(    N^{(S^{(0)})} k n^{(S^{(0)})} t^{(op)}+ t_0 + ({\mbox{depth}}(W_{ER}) + kN^{(S^{(0)})}+ N^{(S^{(0)})} k n^{(S^{(0)})}  
) t^{(op)} \Big)  \lambda^{-2}\right) =\\\nonumber
&&
O\left(\Big(   N^{(S^{(0)})} k n^{(S^{(0)})} t^{(op)} + t_0 + {\mbox{depth}}(W_{ER})  t^{(op)} \Big)  \lambda^{-2}\right).
\end{eqnarray}
}
The appropriate circuit is present  in Fig.\ref{Fig:arb}.
\begin{figure*}[!]
\centering
\includegraphics[scale=0.5]{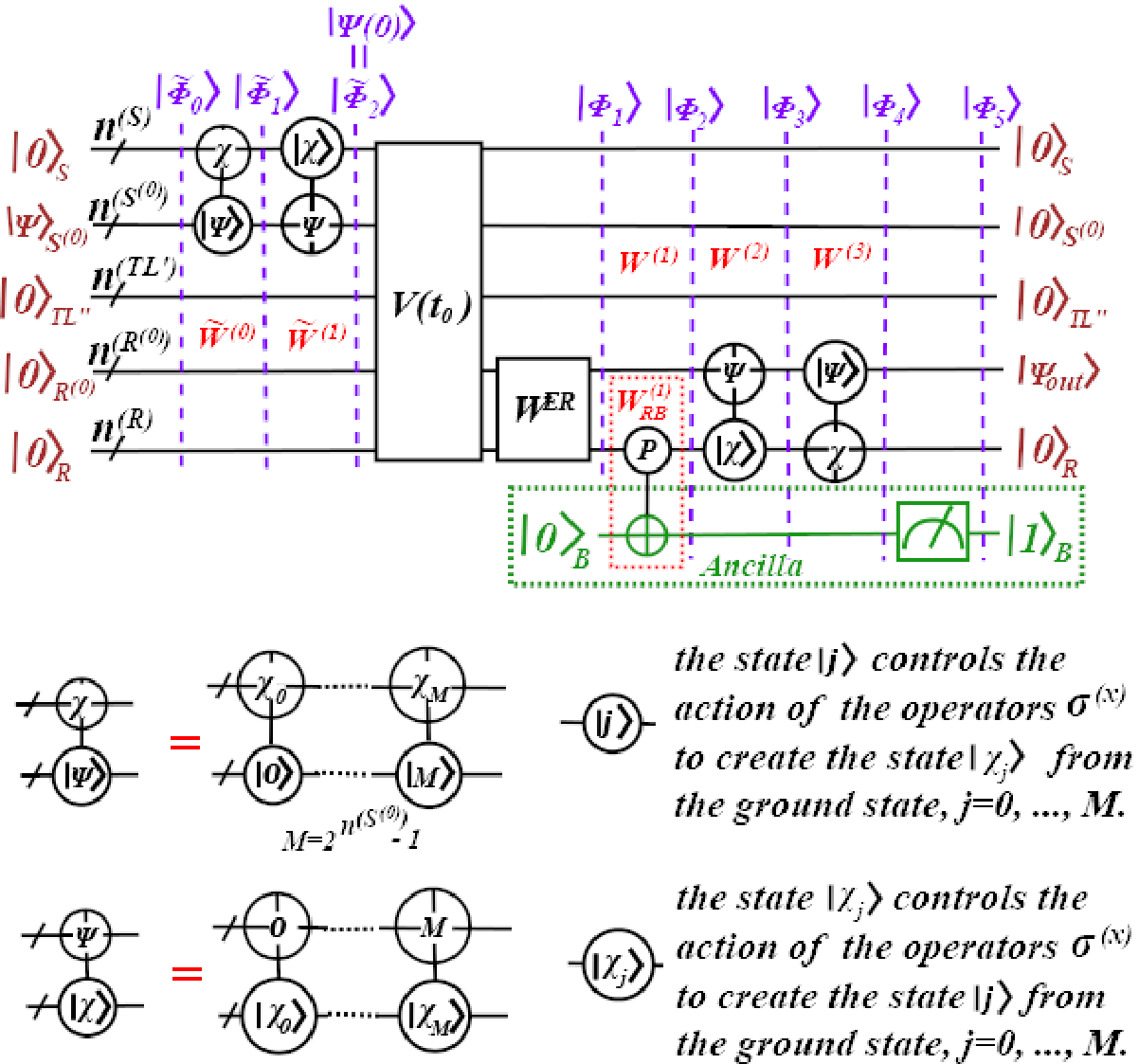} 
\caption{The circuit for the perfect transfer of  an arbitrary    pure state from the subsystem $S^{(0)}$ to the subsystem $R^{(0)}$. The circuit for $W^{(1)}_{RB}$ is shown in Fig. \ref{Fig:1ex}b,c for particular cases. { Here, the state $|\tilde \Phi_0\rangle$ is the initial state of the whole system given in Eq. (\ref{Phi00}); the states $|\tilde \Phi_j\rangle$,   $j=1,2$, given in Eqs.(\ref{tPhi10})  and (\ref{Psi000}) respectively, are the intermediate states performing encoding the  state $|\Psi\rangle_{S^{(0)}}$ (to be transferred to $R^{(0)}$) into the $k$-excitation state of the sender $S$. The states $|\Phi_j\rangle$, $j=1, \dots, 5$, given in Eqs.(\ref{Phi1star2}), (\ref{Phi200}), (\ref{Phi33}), (\ref{Phi42}), (\ref{Phi62}) respectively, are the intermediate states in the state-restoring and state-decoding subroutines. The result appears  in the final state $|\Psi_{out}\rangle$ of the subsystem $R^{(0)}$, see Eq.(\ref{Psiout0}).}}  
\label{Fig:arb}
\end{figure*}
We shall note that the probability amplification considered in Sec.\ref{Section:OM0} is applicable here because the success probability equals $\lambda^2$ and can be bigger than 1/2. 
\section{Restoring of $k$-excitation state}
\label{Section:rest}
According to Sec.\ref{Section:ArbStTr}, the transfer of an arbitrary pure quantum  state can be reduced to the transfer of the appropriate  $k$-excitation state  studied in  Sec. {\ref{Section:FixEx}. Therefore, we consider the state restoring algorithm for the  $k$-excitation state in details.

\subsection{General state-restoring  formulae}
\label{Section:gen}
In this section, we discuss the restoring algorithm for the transferred  $k$ excitation state.

Before fixing the parameters $\varphi$, the receiver  state $|\psi(t_0,\varphi)\rangle_R$ in the transferred  state  $|\Phi_1\rangle$  given in Eq.(\ref{Phi1star2})  has the following structure:
\begin{eqnarray}\label{psiR}
|\psi(t_0,\varphi)\rangle_R = \sum_{j=0}^{N^{(R)}_k-1} r_j(t_0,\varphi) |\chi_j\rangle_R,
\end{eqnarray}
where  {  $|\chi_j\rangle$ represent the basis states in the $k$-excitation receiver state-space,} $r_j$ is a linear combination of the probability amplitudes $s_j$ of the sender initial state, defined in Eq.(\ref{Psik}).
Since the considered dynamics doesn't mix the basis states with different excitation numbers,
we can write 
\begin{eqnarray}\label{Rn}
r_j(t_0,\varphi)  =  \sum_{m=0}^{N^{(S)}_k-1}  T^{(k)}_{jm}(t_0,\varphi) s_m,\;\;j=0,\dots, N^{(S)}_k-1,
\end{eqnarray}
where {some $s_m$ can be zero}, $T^{(k)}$ is the submatrix  of the product $W^{(k)} V^{(k)}$ (details of this submatrix are discussed below in Sec.\ref{Section:W}) and  $W^{(k)}$, $ V^{(k)}$ are  the $k$-excitation blocks of the operators $W$ and $V$, see
Eqs.(\ref{HV}), (\ref{WW2}).
Now we have to solve the system
\begin{eqnarray}\label{T}
T^{(k)}_{jm}(t_0,\varphi) =T^{(k)}_{jj}(t_0,\varphi) \delta_{jm} = \lambda \delta_{jm} ,\;\; j,m= 0,\dots,N^{(S)}_k-1 ,
\end{eqnarray}
where $\delta_{jm}$ is the Kronecker symbol. 
This system consists of two subsystems of equations.  The first one,
\begin{eqnarray}\label{TT1}
T^{(k)}_{jm}(t_0,\varphi) =0,\;\; j\neq m,\;\; j,m= 0,\dots,N^{(S)}_k-1,
\end{eqnarray}
is the restoring system setting zero all unnecessary terms in the expression for $r_j$ given in Eq. (\ref{Rn}).
The second subsystem,
\begin{eqnarray}\label{TT2}
T^{(k)}_{jj}=T^{(k)}_{mm},\;\;0\le j < m \le N^{(R)}_k-1,
\end{eqnarray}
set all $\lambda$-parameters equal to each other. 
Let the set $\varphi_0$ solve system (\ref{TT1}), (\ref{TT2}). 
Then 
we can write
\begin{eqnarray}\label{lT}
\lambda= T^{(k)}_{00}(t_0,\varphi_0),
\end{eqnarray}
and Eq.(\ref{Rn}) gets the following form:
\begin{eqnarray}\label{Rnr}
&&
r_j(t_0,\varphi_0)  =\lambda  s_j , \;\; j= 0,\dots,N^{(S)}_k-1.
\end{eqnarray}
This completes the general algorithm  for restoring  an arbitrary state with operator $W_{ER}$ preserving the excitation number in the system. 

{Although the basic structure of the restoring system is presented in this subsection, to estimate the minimal required dimensionality of the extended receiver and find out   analytical derivation of the maximal parameter $\lambda$ we present detailed analysis of system (\ref{T}) in the next subsection.}

\subsection{General form of operator $W_{ER}$ preserving excitation number}
\label{Section:W}
In this section, we discuss the formulae of Sec.\ref{Section:gen} in more detail. 
The operator $W_{ER}$ serves to restore the transferred state. 
  Due to the fact that we deal with the $k$-excitation state and consider only those terms in the  superposition states $|\Phi_2\rangle$ given in  Eq.(\ref{Phi2}) that collect all excitations in the  receiver $R$, we can select the  block in the product $W V$ (see Eq.\ref{Psi0}) that transfers $|\psi(0)\rangle_S$, Eq.(\ref{Psik}), from  the state $|\Psi(0)\rangle$ given in Eq.(\ref{PsiIn}),  to $|\psi(t_0,\varphi_0)\rangle_R$ in  the state $|\Phi_2\rangle$ given in Eq.(\ref{Phi2}). 
{ For convenience, we use the multi-index notations  with subscripts indicating the appropriate subsystem. { For instance, $i_S$ runs the basis states of the subsystem $S$, $i_{TL}$ runs the basis states of the subsystem $TL$ and so on. Therefore, in the element of $V$,  $V^{(k)}_{ij}$,  we identify the subscripts $i$ and $j$ with the multi-indexes as follows:
$i\equiv \{i_Si_{TL}i_R\}$, $j\equiv \{ j_Sj_{TL}j_R\}$, thus 
  $V_{ij}\equiv V^{(k)}_{i_Si_{TL}i_R; j_Sj_{TL}j_R}$;  in the  element of $W_{ER}$, $(W_{ER})_{nm}$, we  identify the subscripts $n$ and $m$ with the multi-indexes  as follows: $n\equiv n_{ER} \equiv  \{n_An_R\}$,  $m\equiv m_{ER} \equiv  \{m_Am_R\}$, thus  $(W_{ER})_{nm}\equiv (W_{ER})_{n_{ER};m_{ER}}\equiv (W_{ER})_{n_An_{R};m_{ER}}\equiv (W_{ER})_{n_{ER};m_Am_{R}}\equiv (W_{ER})_{n_An_{R};m_Am_{R}} $, because this operator is applied only to the nodes of the extended receiver and $ER=A\cup R$.  At first glance,  multi-indexes look cumbersome, but they are very convenient for enumerating spins in each particular subsystem providing effective numerical simulations of the state-restoring algorithm. }}
In  such notations,  Eq. (\ref{Rn}) can be written as 
\begin{eqnarray}\label{RnI}
r_{j_R}(t_0,\varphi)  =  \sum_{m_S=0}^{N^{(S)}_k-1}  T^{(k)}_{j_Rm_S}(t_0,\varphi) s_{m_S}.
\end{eqnarray}
For  $T^{(k)}_{j_Rm_S}$, we have
\begin{eqnarray}\label{RnT}
&&T^{(k)}_{j_Rm_S} =\sum_{n_{ER}=0}^{N^{(ER)}_k-1} (W_{ER}^{(k)})_{0_Aj_R;n_{ER}} V^{(k)}_{0_S0_{TL'} n_{ER}; m_S0_{TL}0_R}, \\\nonumber
&& j_R = 0,\dots,N^{(S)}_k-1, \;\;  m_S = 0,\dots,N^{(S)}_k-1.
\end{eqnarray}
We recall that $TL = TL'\cup A$; {  $N^{(S)}_k=\binom{n^{(S)}}k$
 and $N^{(ER)}_k=\binom{n^{(ER)}}k$ are the dimensionalities of the $k$-excitation subspaces of, respectively, $S$ and $ER$.}
Introducing matrices $\hat W_{ER}$ and $\hat V$ with the elements
$(\hat W_{ER})_{j_R;n_{ER}}  =(W_{ER}^{(k)})_{0_Aj_R;n_{ER}}$ and $\hat V_{n_{ER};m_S} = V^{(k)}_{0_S0_{TL'} n_{ER}; m_S0_{TL}0_R}$, we can write 
$\vec r = \hat W_{ER} \hat V \vec s$, where $\vec r$ and $\vec s$ are  the column-vectors with entries, respectively, $r_{j_{R}}$ and $s_{m_S}$. {Here the index $0_P$ corresponds to the ground state of the subsystem $P$.}
Since $\hat W_{ER}$ is the block of the unitary matrix $W_{ER}$, it inherits the orthonormality condition
\begin{eqnarray}\label{Wort}
\hat W_{ER} \hat W_{ER}^\dagger = I_{R},
\end{eqnarray}
where $I_R$ is the identity matrix acting in the state space of $R$. 
The restoring constraints in Eq.(\ref{TT1}) and constraints on the $\lambda$-parameters in Eq.(\ref{TT2}) yield another constraint on $\hat W_{ER}$:
\begin{eqnarray}\label{Wort2}
 &&\hat W_{ER} \hat V =\lambda I_R, \;\; \lambda =  (\hat W_{ER} \hat V)_{00} 
 \end{eqnarray}
 { This constraint can be split into two following systems:}
 \begin{eqnarray}
 \label{Wort31}
 &&  (\hat W_{ER} \hat V)_{i_R j_R}=  0,\;\; i_R \neq  j_R,\;\;i_R, j_R=0,\dots N^{(S)}_k-1,\\\label{Wort32}
 &&
   (\hat W_{ER} \hat V)_{00}=   (\hat W_{ER} \hat V)_{i_R i_R},\;\;i_R=1,\dots N^{(S)}_k-1.
\end{eqnarray}
The {restoring and} optimizing parameters are the elements of $\hat W_{ER}$:
\begin{eqnarray}\label{varphi}
\varphi &=&\{ {\mbox{Re}}(\hat W_{ER})_{j_Rn_{ER}}, {\mbox{Im}}(\hat W_{ER})_{j_Rn_{ER}}:\\\nonumber
&& j_R=0,\dots, N^{(S)}_k-1,\;  n_{ER}=0,\dots, N^{(ER)}_k-1\}.
\end{eqnarray}
 They must satisfy  orthonormality condition   (\ref{Wort}) and restoring constraints (\ref{Wort31}), (\ref{Wort32})  and maximize the  parameter $\lambda$. 

{  \subsubsection{Required dimensionality of extended receiver}}
\label{Section:DimER}
First of all, 
{let us define the number of free parameters in the operator $\hat W_{ER}$ 
(i.e., the number of parameters $\varphi$ in  Eq.(\ref{varphi}) that remain free after satisfying the orthogonality constraint (\ref{Wort}))
and compare it with the number of equations in the restoring system (\ref{Wort31}), (\ref{Wort32}). There are $2 N^{(ER)}_k N^{(S)}_k$ real parameters  in $\hat W_{ER}$. They are subjected to the orthonormalization  condition (\ref{Wort})
 which yields $(N^{(R)}_k)^2$ equations thus leaving $N^{(par)} = 2 N^{(ER)}_k N^{(R)}_k - (N^{(R)}_k)^2 = N^{(R)}_k (2  N^{(ER)}_k - N^{(R)}_k)$ free  parameters in   $\hat W_{ER}$. 
 These parameters are used to satisfy $N^{(eq)}_1= 2N^{(R)}_k (N^{(R)}_k-1)$ real restoring equations   (\ref{Wort31}) and $N^{(eq)}_2=2(N^{(R)}_k-1)$ real conditions (\ref{Wort32}). Therefore, for successful restoring with equal $\lambda$-parameters we have to satisfy inequality 
 \begin{eqnarray}\label{ne}
 &&
 N^{(par)}\ge N^{(eq)}  =N^{(eq)}_1+N^{(eq)}_2   \;\;\;\Rightarrow \\\nonumber
 && N^{(ER)}_k \ge N^{(ER;v)}_k\equiv \frac{3}{2} N^{(R)}_k - \frac{1}{ N^{(R)}_k},\;\; N^{(R)}_k=N^{(S)}_k.
 \end{eqnarray}
This is the  necessary number of $k$-excitation  states in the extended receiver. 

{However, there is another requirement   to   the dimensionality of the extended receiver that increases the required $N^{(ER)}_k$. This requirement is  related to the orthogonality of the columns of $W_{ER}$ and $\hat V$ and 
is formulated  in the following Proposition.

{\bf Proposition 1.} Dimensionality of the $k$-excitation spaces of the extended receiver and sender are related by the following inequality:
\begin{eqnarray}\label{NERmin}
N^{(ER)}_k\ge N^{(ER;min)}_k \equiv 2 N^{(S)}_k-1.
\end{eqnarray} 

{\it Proof.}
  { Let us denote  the rows of the matrix $W_{ER}^{(k)}$  and  the columns of the matrix $\hat V$  by, respectively,  $a_j$, $j=0,\dots,N^{(ER)}_k-1$,  and  $b_j$, $j=0,\dots,N^{(S)}_k-1$.}
The system (\ref{Wort2}) can be written as
\begin{eqnarray}\label{ab}
a_i b_j=\lambda \delta_{ij},\;\; i, j=0,\dots,N^{(S)}_k-1.
\end{eqnarray}
 It prescribes the following expansion for $b_j$:
\begin{eqnarray}\label{rel1}
b_j = \lambda a_j^\dagger + \sum_{k=N^{(S)}_k}^{N^{(ER)}_k-1}  \alpha_{jk} a_k^\dagger ,\;\; j=0,\dots,N^{(S)}_k-1,\;\; \alpha_{jk}=a_k b_j,\;\; \lambda = a_i b_i,\;\; \forall i.
\end{eqnarray}
If inequality (\ref{NERmin}) is violated, i.e., 
\begin{eqnarray}\label{Neq}
N^{(ER)}_k< 2 N^{(S)}_k-1,
\end{eqnarray} 
then all $a_j^\dagger$, $j=N^{(S)}_k,\dots, N^{(ER)}_k-1$ can be expressed in terms of $(b_i-\lambda a_i^\dagger)$, $i=0,\dots, N^{(ER)}_k-N^{(S)}_k-1$ using $(N^{(ER)}_k-N^{(S)}_k)$ equations from system (\ref{rel1}):
\begin{eqnarray}
a_j^\dagger=\sum_{i=0}^{N^{(ER)}-N^{(S)}-1} \gamma_{ji} (b_i-\lambda a_i^\dagger),\;\; j=N^{(S)}_k,\dots, N^{(ER)}_k-1,
\end{eqnarray}
{where $\gamma_{ji}$ are some coefficient depending on $\alpha_{lk}$.}
Then the rest  $N^{(S)}_k-(N^{(ER)}_k-N^{(S)}_k)= 2N^{(S)}_k-N^{(ER)}_k\stackrel{(\ref{Neq})}{\ge} 2$  equations  from system (\ref{rel1}) form the system of  linear equations relating the vectors $(b_i-\lambda a_i^\dagger)$, $i=0,\dots N^{(S)}_k-1$:
\begin{eqnarray}\label{rel2}
&&
b_k-\lambda a_k^\dagger =\sum_{i=0}^{N^{(ER)}_k-N^{(S)}_k-1} \Gamma_{ki}  (b_i-\lambda a_i^\dagger),\;\; k= N^{(ER)}_k-N^{(S)}_k,\dots, N^{(S)}_k-1, \\\nonumber
&&\Gamma_{ki}=  \sum_{l=N^{(S)}_k}^{N^{(ER)}_k-1}  \alpha_{kl}  \gamma_{li}.
\end{eqnarray}
Multiplying Eq.(\ref{rel2}) by $b_j^\dagger $, $0\le j\le N^{(S)}_k-1$, from the left yields
\begin{eqnarray}\label{bb}
&&
b^\dagger_j b_k - \delta_{kj} \lambda^2= \sum_{i=0}^{N^{(ER)}_k-N^{(S)}_k-1} \Gamma_{ki} (b^\dagger_jb_i   - \delta_{ij} \lambda^2),\\\nonumber
&& j=0,\dots, N^{(S)}_k-1,\;\; k= N^{(ER)}_k-N^{(S)}_k,\dots, N^{(S)}_k-1.
\end{eqnarray}
This is a system of $(2N^{(S)}_k-N^{(ER)}_k)N^{(S)}_k $ 
complex  equations and can be considered as a system of equations for $(2 N^{(S)}_k-N^{(ER)}_k) (N^{(ER)}_k-N^{(S)}_k) \le  (2 N^{(S)}_k-N^{(ER)}_k)(N^{(S)}_k-2)$ complex  parameters $\Gamma_{ki}$ and  one real  parameter  $\lambda^2$. Therefore, the number of equations exceeds the number of parameters and, consequently,  the above system (\ref{bb})  generates additional constraints for the entries of the vectors $b_i$ which are not presumed by the state transfer process. 

{On the contrary, if  inequality (\ref{NERmin}) holds, then system (\ref{rel2})  either is empty  if  $N^{(ER)}_k> 2 N^{(S)}_k -1$, or includes single  equation if $N^{(ER)}_k= 2 N^{(S)}_k -1$. In the latter case,  system (\ref{bb}) consists of $N^{(S)}_k$ equations for $N^{(S)}_k-1$ $\Gamma$-parameters and $\lambda^2$.  It is important that, after eliminating the $\Gamma$-parameters, system (\ref{bb}) reduces to the polynomial equation for $\lambda^2$ with real coefficients, see Proposition 2 below. 
 Therefore,  there is no additional  constraints for the entries of the vectors $b_i$ in both cases.}$\Box$

Thus, if inequality (\ref{NERmin})  holds, then the  transformation $\hat W_{ER}$ with imposed normalization condition (\ref{Wort})  includes enough free parameters to satisfy the restoring system  (\ref{Wort31}), (\ref{Wort32})  because
\begin{eqnarray}\label{NERneq}
N^{(ER;min)}_k > N^{(ER;v)}_k, \;\;\;\forall N^{(S)}_k.
\end{eqnarray}}
The parameter  $N^{(ER;min)}_k$ uniquely determines the required minimal number of qubits $n^{(ER)}$ in the extended receiver. However, to obtain the large enough $\lambda$ we have to use significantly larger extended receiver, which is  confirmed in  examples of Sec.\ref{Section:num}.} 

{It follows from the above consideration that the number of free parameters in the transformation $W_{ER}$ exceeds the number of restoring equations (\ref{Wort31}), (\ref{Wort32}).  The extra parameters serve to enlarge the 
absolute value of  $\lambda$. The preferable way to reach this goal is to put to zero  as many probability amplitudes  as possible in the garbage $|g_1\rangle$ in $|\Phi_1\rangle$, defined in Eq. (\ref{Phi1star2}), using these extra parameters.
Namely,  along with constraints (\ref{ab})  (or (\ref{Wort2})), we involve the following $N^{(ad)}_k N^{(S)}_k$  complex constraints, $N^{(ad)}_k = N^{(ER)}_k -N^{(ER;min)}_k$:
\begin{eqnarray}\label{ab2}
a_i b_{j_R}=0,\;\;i=N_k^{(S)},  \dots,N_k^{(S)}+N^{(ad)}_k-1  =
N^{(ER)}_k-N^{(S)}_k , \;\;  j_R=0,\dots,N^{(S)}_k-1,
\end{eqnarray}
where $i$ enumerates the basis states of the extended receiver. Thus, we generalize Eq.(\ref{Wort31}) replacing $\hat W$ with $W^{(k)}_{ER}$ as follows:
\begin{eqnarray}
 \label{Wort312}
(W^{(k)}_{ER} \hat V)_{i j_R}=  0,\;\; i \neq  j_R, \;\;  j_R=0,\dots,N^{(S)}_k-1,\;\; i=0,\dots,  N^{(ER)}_k-N^{(S)}_k
\end{eqnarray}
and  replace  orthogonality condition (\ref{Wort}) with the following one:
\begin{eqnarray} \label{Wort313}
(W_{ER}^{(k)} (W_{ER}^{(k)})^\dagger)_{i j}=  \delta_{ij}, \;\;  i,j=0,\dots,  N^{(ER)}_k-N^{(S)}_k.
\end{eqnarray}
Now the list of restoring and optimizing parameters is following:
\begin{eqnarray}\label{varphi3}
\varphi &=&\{ {\mbox{Re}}( W_{ER})_{j n_{ER}}, {\mbox{Im}}(W_{ER})_{jn_{ER}}:\\\nonumber
&& j=0,\dots, N^{(ER)}_k- N^{(S)}_k,\;  n_{ER}=0,\dots, N^{(ER)}_k-1\}.
\end{eqnarray}
Obviously, constraints (\ref{Wort31}) are included in Eq.(\ref{Wort312}). We note that Eq.(\ref{Wort313}) is not the complete unitarity condition for $W_{ER}$ because the unitarity condition requires  $i,j=0,\dots,  N^{(ER)}_k-1$.
The number of free real parameters of the unitary transformation $W^{(k)}_{ER}$  involved in conditions (\ref{Wort312}) and constrained by  $(N^{(ER)}_k-N^{(S)}_k+1)^2$ real conditions (\ref{Wort313}) is 
$2 N^{(ER)}_k (N^{(ER)}_k-N^{(S)}_k+1) - (N^{(ER)}_k-N^{(S)}_k+1)^2 =(N^{(ER)}_k)^2-(N^{(S)}_k -1)^2 $.  By virtue of condition (\ref{NERmin}), this  is enough to satisfy all  $N^{(eq)}+2 N^{(ad)}_k N^{(S)}_k =2 (N^{(ER)}_k N^{(S)}_k + N^{(S)}_k - (N^{(S)}_k)^2-1)$ real constraints (\ref{Wort32}) and (\ref{Wort312}).  
It is interesting that restoring constraints  (\ref{Wort32}),  (\ref{Wort312}) and orthogonality  condition (\ref{Wort313})  allow to avoid further maximization of $\lambda$, although they do not uniquely  fix all free parameters of the unitary transformation $W_{ER}$.  
More exactly, the following proposition holds.

{\bf Proposition 2.}     If  restoring constraints (\ref{Wort32}),  (\ref{Wort312}) and orthogonality  condition (\ref{Wort313})  are satisfied,
then   $\lambda^2$  is {   among the eigenvalues  of the Hermitian Gram matrix  $\hat V^\dagger \hat V$ and therefore is nonnegative real number. It can be found as a root of the characteristic equation 
\begin{eqnarray}\label{lamDp}
{\mbox{det}}(\hat V^\dagger \hat V - \lambda^2 I) =0, 
\end{eqnarray}
(where $I$ is the identity matrix) which is the
$N^{(S)}_k$-order polynomial   equation  with real coefficients depending only on the elements of $\hat V$, which, in turn,  depend on the Hamiltonian, dimension of the extended receiver $n^{(ER)}$ and the time instant for state registration $t_0$.}


{\it Proof.}  Instead of Eq.(\ref{rel1}), we have the following expansion for the vectors $b_j$, satisfying constraints (\ref{Wort32}) and (\ref{Wort312}) :   
\begin{eqnarray}\label{rel11}
b_j =\lambda a_j^\dagger+ \sum_{k=N^{(ER)}_k-N^{(S)}_k+1}^{N^{(ER)}_k-1}  \alpha_{jk} a_k^\dagger ,\;\; j=0,\dots,N^{(S)}_k-1.
\end{eqnarray}
Eliminating  {  $N^{(S)}_k-1$ elements} $a_k^\dagger$,  $N^{(ER)}_k-N^{(S)}_k+1\le k\le N^{(ER)}_k-1$, from {  system of $N^{(S)}_k$ equations} (\ref{rel11})  we end up with the single relation 
\begin{eqnarray}\label{rel22}
&&
\sum_{i=0}^{N^{(S)}_k-2} \Gamma_{i}  (b_i-\lambda a_i^\dagger)=b_{N^{(S)}_k-1}-\lambda a_{N^{(S)}_k-1}^\dagger 
\end{eqnarray}
with some coefficients $\Gamma_i$ { expressed in terms of $\alpha_{jk}$ (we do not need the exact expressions for $\Gamma_i$).}
{  Multiplying  Eq.(\ref{rel22}) by $b_j^\dagger$, $0\le j\le N^{(S)}_k-1$, from the left  yields the system of $N^{(S)}_k$ scalar complex equations, which includes only vectors $b_j$, $ j=0,\dots,N^{(S)}_k-1$, (i.e., the elements of $\hat V$) and do not depend on the elements of $W_{ER}$:
\begin{eqnarray}\label{bb1}
 \sum_{i=0}^{N^{(S)}_k-2} \Gamma_{i} (b^\dagger_j  b_i - \delta_{ij} \lambda^2) =b^\dagger_j b_{N^{(S)}_k-1} - \delta_{j,N^{(S)}_k-1} \lambda^2 ,\;\; j=0,\dots,N^{(S)}_k-1 ,
\end{eqnarray}
Eliminating $\Gamma_{i}$ from this system we end up with the single $N^{(S)}_k$-order polynomial  equation  for $\lambda^2$ with real coefficients depending on the elements of $\hat V$.  {   This equation can be written as the characteristic equation for the Gram matrix, Eq.(\ref{lamDp}). {   Positivity of $\lambda^2$ follows from the nonnegative definiteness of the Gram matrix $G$   (zero $\lambda^2$ is out of our interest). }}
  $\Box$
  
 We will derive and solve  the  polynomial 
 equation { (\ref{bb1})} for $\lambda$  in Sec.\ref{Section:num}  for the particular examples, see Eq.(\ref{bb3}). 
 
 We shall note that Proposition 2 { is stipulated by} condition (\ref{Wort32}) that equate all $\lambda$-parameters.  It is not applicable to the restoring algorithms  in Refs.\cite{FPZ_2021} where all $\lambda$-parameters are different. 
 
 Thus, the absolute value of $\lambda$ is completely defined by the entries of the matrix $\hat V$. This matrix depends on the time $t_0$, while its dimensionality is defined by the dimensionality of the extended receiver. Consequently, $\lambda$ also depends on  $t_0$ and $N^{(ER)}_k$. Proposition 2 removes necessity to solve {  the maximization  problem for $\lambda$} in the full extend, because any set of parameters $\varphi$ satisfying the restoring system (\ref{Wort32}), (\ref{Wort312}), (\ref{Wort313}) yields  $\lambda$ from the list of roots of the polynomial equation (\ref{lamDp}). {  Nevertheless, we have to solve the restoring system, at least, for two following purposes.  First,  solving the restoring system we construct  the unitary transformation $W_{ER}$, {which is required for realizing the state restoring in the PST- algorithm.}  Second,  to  clarify  which of the supplied roots of polynomial equation (\ref{lamDp})  is realizable by the unitary transformation. In the examples of Sec.\ref{Section:num}, this is the minimal root. At the moment, we can not be sure that $W_{ER}$ can yield only the minimal root $\lambda$ of the polynomial equation  (\ref{lamDp}) for any dimension of sender and any excitation number $k$. }
 }

\subsection{Particular realization of operator $W_{ER}$ preserving excitation number}
\label{Section:WPr}

{ The operator $W_{ER}$ of general form can be realized using, for instance, the algorithms based on the Gray code \cite{BBCDMSSSW,VMS}. }
In this section we consider a particular block-wise  realization \cite{WSW} of the excitation-preserving operator $W_{ER}$ based on the excitation preserving two-qubit  operators $U_{ij}(\alpha,\beta,\gamma)$ applied  to the  $i$th and $j$th qubits \cite{DFZ_2020}:
\begin{eqnarray}\label{Uij}
U_{ij}(\alpha,\beta,\gamma) ={ e^{i \gamma}} C_{ij} R_i(\alpha,\beta) C_{ji} R_i^\dagger(\alpha,\beta) C_{ij},
\end{eqnarray}
where $C_{ij}$ is CNOT,
\begin{eqnarray}
C_{ij} = |1\rangle_i\, {_i\langle 1|}\otimes \sigma^{(x)}_j +  |0\rangle_i\, {_i\langle 0|}\otimes I_j,
\end{eqnarray}
$\sigma^{(x)}_j$ and  $I_j$ are, respectively, the operator $\sigma^{(x)}$ and identity operator applied to the $j$th qubit of the extended receiver, and
 \begin{eqnarray}
 R_i(\alpha_i,\beta_i)= R_{zi}(\beta_i) R_{yi}(\alpha_i) R^\dagger_{zi}(\beta_i).
 \end{eqnarray}
Here, $R_{zi}$ and $R_{yi}$ are $z$- and $y$-rotation of the $i$th qubit of the extended receiver:
\begin{eqnarray}\label{yz}
R_z(\alpha)=e^{-\frac{i \sigma_z \alpha}{2}},\;\; R_y(\beta)=e^{-\frac{i \sigma_y \beta}{2}}.
\end{eqnarray}
Based on these operators, we construct $W_{ER}$, see Fig.\ref{Fig:W0}:
\begin{eqnarray}\label{WV}
W_{ER} =  V_Q\dots V_1,
\end{eqnarray}
where
\begin{eqnarray}\label{defV}
V_k=\prod_{j=1}^{n^{(ER)}}   U_{j, (j+1\!\!\!\mod{n^{(ER)})}} (\alpha_{kj},\beta_{kj},\gamma_{kj}), \;\;k=1,\dots,Q.
\end{eqnarray}
There are $2 Q n^{(ER)}$ real parameters in $W_{ER}$.  In this case, the list of parameters is 
\begin{eqnarray}\label{varphi1}
\varphi = \{ \alpha_{kj},\beta_{kj},  \gamma_{kj}: j=1,\dots,n^{(ER)},\; k=1,\dots, Q\}.
\end{eqnarray}
  { The depth  of $W_{ER}$ is $O(Qn^{(ER)})$, therefore Eq.(\ref{dQ})  now reads:
\begin{eqnarray}\label{qQ1}
&&T^{(PST)}=O\Big((   N^{(S^{(0)})} k n^{(s_0)} t^{(op)} + t_0 +Qn^{(ER)} t^{(op)}) \lambda^{-2}\Big).
\end{eqnarray}}

\begin{figure*}[!]
\centering
\includegraphics[scale=0.4]{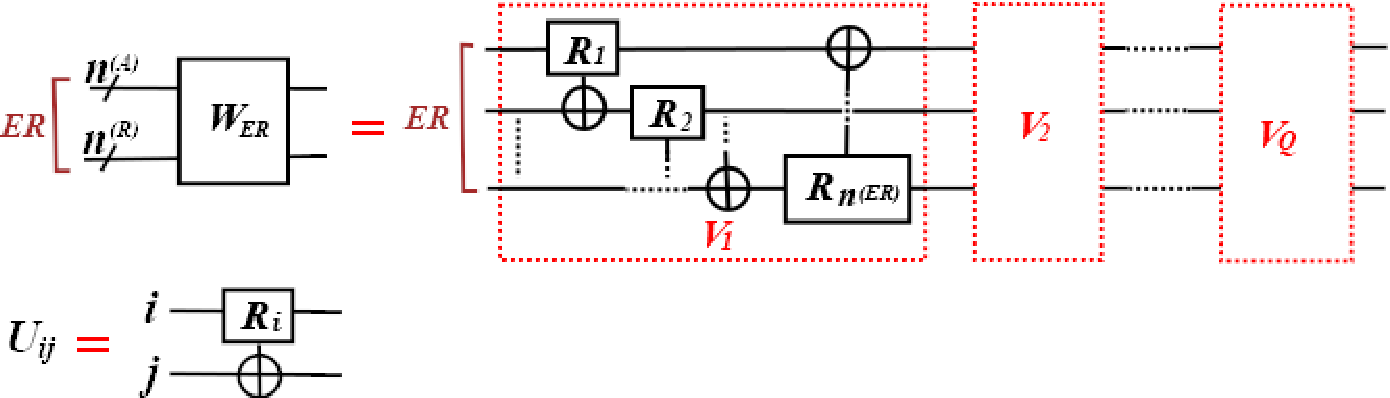} 
\caption{A particular realization of the  unitary transformation $W_{ER}$ which  preserves the excitation number and
is  represented by Eq.(\ref{WV}). {  The blocks $V_j$ are defined in Eq.(\ref{defV}) with the operators $U_{ij}$ given in Eq.(\ref{Uij}).} }  
\label{Fig:W0}
\end{figure*}

{

\section{Numerical   restoring of 3-qubit  2-excitation state}
\label{Section:num}
Thus, the PST of an arbitrary state includes five subroutines described in 
Sec.\ref{Section:ArbStTr}. 
We present the numerical simulation  of the  state-restoring algorithm for the perfect $k$-excitation state transfer, whose theoretical consideration is given  in  Sec.\ref{Section:rest}.

For numerical simulations,  we choose the $XX$-Hamiltonian with all-node dipole-dipole interactions and external magnetic field directed  perpendicular to  the line chain:
\begin{eqnarray}\label{Hamiltonian}
H=\sum_{j>i}D_{ij} (I_{xi}I_{xj} +I_{yi}I_{yj} ), \;\;[H,I_z]=0,
\end{eqnarray}
where $I_{\alpha i}=\frac{1}{2}\sigma^{(\alpha)}$ is the operator of 
the $\alpha$-projection of the $i$th spin momentum, $\alpha=x, y, z$, $I_z=\sum_i I_{zi}$,  $\sigma^{(\alpha)}$  are the Pauli matrices,
$D_{ij}= \gamma^2\hbar/(2 r_{ij}^3)$ is the dipole-dipole coupling constant between the $i$th and $j$th spins,  $r_{ij}$ is the distance between the $i$th and $j$th spins, 
$\gamma$ is the gyromagnetic ratio,  $\hbar$ is the Planck constant  ($\hbar =1$ for simplicity). In simulations, we usually use the homogeneous chains (so that all nearest-neighbor coupling constants equal to each other) and  the {  dimensionless time $\tau = t D_{12}$.}
{First,} we discuss the simulation of  the 2-excitation state restoring in the chain of $N=10$ nodes  with the 3-qubit sender and receiver  ($n^{(S)}=n^{(R)}=N^{(S)}_2=N^{(R)}_2=3$) and different sizes of the external receiver, $n^{(ER)}=4,5,6$.  
Thus, we use the basis in the two-excitation state-subspace of the extended receiver ordered as follows:
\begin{eqnarray}
|1_l1_m\rangle,\;\; l=1,\dots, n^{(ER)}-1,\;\;  m=l+1,\dots, n^{(ER)}
\end{eqnarray}
($1_j$ means that the $j$th spin of the extended receiver is excited) and take into account that $N^{(ER;min)}_2= 2 N^{(S)}_2-1=5$ for our case.  
The result of restoring is characterized by the parameter $\lambda$ and the  time instant for state registration $\tau_0$. 
  To find these parameters  we proceed as follows.  

According to Proposition 2 in Sec. \ref{Section:DimER}, { $\lambda^2$} is among solutions of the polynomial equation (\ref{lamDp}) reduced from system (\ref{bb1}).
In our example this system  reads
\begin{eqnarray}\label{bb2}
 \sum_{i=0}^{1} \Gamma_{i} (  b^\dagger_j b_i  - \delta_{ij} \lambda^2)=b^\dagger_j b_{2} - \delta_{j2} \lambda^2  ,\;\; j=0,1,2,
\end{eqnarray}
and, { after eliminating $\Gamma_i$, $i=0,1$, it reduces to the cubic  polynomial equation (\ref{lamDp})   for $\lambda^2$:}
\begin{eqnarray}\label{bb3}
&&
\lambda^6 - (b_0^\dagger b_0 +b_1^\dagger b_1+ b_2^\dagger b_2) \lambda^4 +\\\nonumber
&&( b_0^\dagger b_0 b_1^\dagger b_1+ b_0^\dagger b_0 b_2^\dagger b_2 +  b_1^\dagger b_1 b_2^\dagger b_2 -
|b_0^\dagger b_1|^2 -|b_0^\dagger b_2|^2 -|b_1^\dagger b_2|^2 )\lambda^2 +\\\nonumber
&& |b_0^\dagger b_2|^2  b_1^\dagger b_1+ |b_1^\dagger b_2|^2  b_0^\dagger b_0+ |b_0^\dagger b_1|^2  b_2^\dagger b_2 - 
b_0^\dagger b_0b_1^\dagger b_1b_2^\dagger b_2  - {   2}\, {\mbox{Re}}(b_1^\dagger b_0 b_2^\dagger b_1 b_0^\dagger b_2)=0.
\end{eqnarray}
{  Eq. (\ref{bb3}) justifies the   statement in Proposition 2 that} the solution 
 $\lambda^2$ of  this equation does not depend on the parameters of the unitary transformation  $W_{ER}$, { but depends on the dimension of the  extended receiver $n^{(ER)}$  and the time instant for state registration $\tau_0$ because vectors $b_j$ depend on those parameters.}  {The question remains which of three roots can be realized by the unitary transformation $W_{ER}$. The preliminary study shows the realizability of the smallest root $\lambda^2$ {   (i.e., the smallest eigenvalue of the Gram matrix $G$)} trough the solution of the restoring system, while the bigger roots have never been obtained.} {The conditions for realizability of the bigger roots are not clear yet. Perhaps, there is a principal obstacle for realizing them.} Therefore, to find the time instant for state registration,  we fix $n^{(ER)}$, numerically construct the  $\tau$-dependence of the minimal root $\lambda_{min}^2$  of Eq.(\ref{bb3}) with the time-step $\Delta\tau=0.001$ and find its maximum over  $\tau$ together with appropriate time instant $\tau_0$.
The $\tau$-dependence of $\lambda_{min}^2$ for $n^{(ER)}=4,5,6$ is present  in Fig.\ref{Fig:LT} over $0\le \tau\le 20$. The maximal values of $\lambda_{min}^2$ together with the appropriate time instances $\tau_0$ are marked in this figure. It is important that the graphs are bell shaped in the neighborhood of $\tau_0$. This means that a small deviation from the  predicted $\tau_0$ does not significantly decrease $\lambda^2$.
\begin{figure*}[!]
\centering
\includegraphics[scale=0.7]{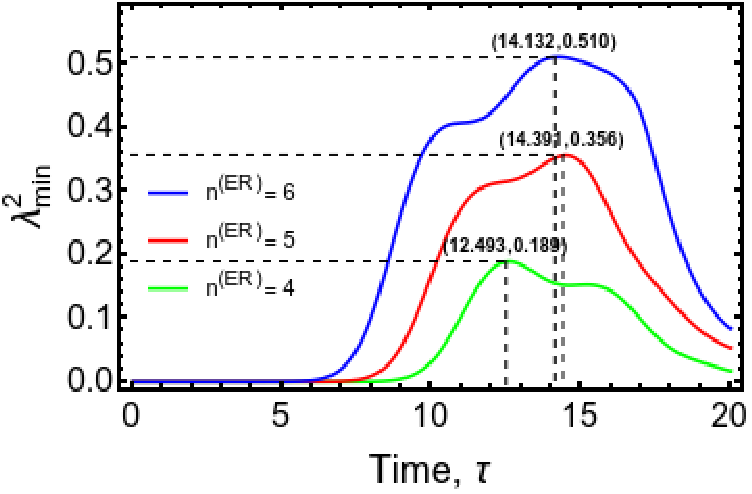} 
\caption{The time-dependence of the absolute value of the scale factor $\lambda_{min}^2$ { (the minimal root of the polynomial  equation (\ref{bb3}))} for different numbers of qubits in the extended receiver  $n^{(ER)}=4,5,6$. The maximal values of $\lambda_{min}^2$ together with the appropriate time instances $\tau_0$ are marked. { Here $\tau=D_{12} t$ is the dimensionless time. This figure demonstrates that the minimal root $\lambda_{\min}^2$  increases with an increase in $n^{(ER)}$.} }  
\label{Fig:LT}
\end{figure*}

Next, at the found time instant $\tau_0$, we solve the restoring system (\ref{Wort32}),  (\ref{Wort312}),  (\ref{Wort313}) and confirm that the  found  $\lambda^2$  coincides with   $\lambda_{min}^2(\tau_0)$, predicted by solving Eq. (\ref{bb3}). Results of these simulations are collected in Table  \ref{Table:T1}a,  where the success probability $\lambda^2$ is also indicated. {   This table shows that $\lambda^2$ increases with an increase in $n^{(ER)}$. 

In addition, $\lambda^2$ can be increased even more if we replace the homogeneous spin chain with the high-fidelity spin chain introducing the weak end-bonds as demonstrated below for the longer spin-chain of 42 nodes in Table \ref{Table:T2}}. {We emphasize that, although the solution of the restoring system is not unique, all the solutions lead to the same $\lambda^2_{min}$. }

 Below we present all three roots of Eq. (\ref{bb3}) for each considered $n^{(ER)}$ at appropriate time instant $\tau_0$ given in Table \ref{Table:T1}a:
 \begin{eqnarray}\label{lll}
&&\lambda = 0.435,\; 0.660, \;0.828, \;\;\;  n^{(ER)}=4,\\\nonumber
&&
\lambda= 0.597, \;0.794, \;0.866,\;\;\;n^{(ER)}=5,\\\nonumber
&&
\lambda= 0.714, \;0.888, \;0.931,\;\;\; n^{(ER)}=6.
\end{eqnarray}

Now we turn to the particular realization of $W_{ER}$, proposed in Sec.\ref{Section:WPr},  {for $Q=2,3$}.
Again we use the restoring system   (\ref{Wort32}),  (\ref{Wort312}),  (\ref{Wort313}).
{   However, {now}   we do not have enough free parameters $\varphi$, collected  in list (\ref{varphi1}),  in the unitary transformation $W_{ER}$ to satisfy all equations (\ref{Wort312}).  Therefore, we replace the parameter $N^{(ad)}_2$ with $\tilde N^{(ad)}_2{ <N^{(ad)}_2 }$, thus the {   subscript $i$  in  Eq.(\ref{Wort312})  takes the values}  $i=1,\dots, N^{(S)}_2 +\tilde N^{(ad)}_2$. } In this case the parameter $\lambda$ depend on the elements of the  unitary transformation $W_{ER}$, which agrees with Proposition 2. To maximize $\lambda$, we perform the maximization over 1000 solutions of the restoring system.  
The results are collected in Table \ref{Table:T1}b.  
This Table shows  that  $\lambda$ increases with $Q$ at fixed $n^{(ER)}$, while $\lambda$ for $n^{(ER)}=6$ is less than for $n^{(ER)}=4, 5$. The latter  happens because  $\tilde N^{(ad)}_2$ is significantly less then    $ N^{(ad)}_2$ in this case for both $Q=2$ and $3$, therefore we put to zero small part of the probability amplitudes in the garbage.  This disadvantage must disappear with an increase in $Q$, but larger $Q$ is  not considered in this paper.
\begin{table}
\begin{tabular}{|c|ccc|}
\hline
$n^{(ER)}$ & 4 & 5 &6 \cr
$N^{(ER)}_2$ & 6 & 10 &15\cr
$N^{(ad)}_2$ & 1 & 5 &10 \cr
$\tau_0$ &12.493&14.391&14.132\cr
$\lambda$ &$0.435 $ &$0.597$&$ 0.714 $ \cr
$\lambda^2$ &0.189 &0.356&{  0.510}\cr
\hline
\end{tabular}
\\~\\
(a)\\~\\
\begin{tabular}{|c|cccccc|}
\hline
$n^{(ER)}$ & 4 & 4 & 5&5&6&6 \cr
$Q$            &2  &3  &2&3&2&3\cr
$N^{(ER)}_2$ & 6&6 & 10 &10&15&15\cr
$N^{(ad)}_2$&1 &1&5&5&10&10 \cr
$\tilde N^{(ad)}_2$&0 &0&1&2&1&3 \cr
$\tau_0$ &12.493&12.493&14.391&14.391&14.132&14.132\cr
$\lambda$ & $0.434 $& $0.435 $& $0.494$& $0.522$&$0.386$&$0.492$\cr
$\lambda^2$ & 0.188 &0.189  &0.244& 0.272&0.149 &0.242\cr
\hline
\end{tabular}
\\~ \\
 (b)
\caption{Parameters of the $2$-excitation  PST-algorithm  along the $N=10$-node chain   with the three-qubit sender (receiver), $N^{(S)}_2= 3$, using  
 (a) the unitary transformation $W_{ER}$ of general form preserving the excitation number, Sec.\ref{Section:W}; { this table demonstrates an increase in $\lambda^2$ with an increase in the dimensionality of the extended receiver $n^{(ER)}$,}
 and  (b) the unitary transformation   $W_{ER}$  of the particular form presented in  Sec.\ref{Section:WPr}; { $\lambda^2$ increases with $Q$, but is not monotonic in $n^{(ER)}$ for the reason clarified in the text.}
  } 
\label{Table:T1}
\end{table}

We also consider the perfect transfer of  an arbitrary  2-excitation 3-qubit pure state along the longer homogeneous  chains of {   $N=20, 25, 30, 35, 40, 45, 50, 55$} nodes with the 5-node extended receiver $n^{(ER)}=5$, $N^{(ER)}_2=10$ and $N^{(ad)}_2=5$. We  scan the time interval with the same step $0.001$ and study the $\tau$-dependence of the minimal root $\lambda_{min}$ of Eq.(\ref{bb3}). 
In addition, we consider the chain of 42 nodes {   prepared for the high-fidelity state transfer (HFST).} All node interactions are used  in this chain (like in other examples)   and two pairs of  coupling constants  $D_{1,2}=D_{41,42}=0.354\delta$,  $D_{2,3}=D_{40,41} = 0.497\delta$, adjusted for maximizing $\lambda^2$ \cite{ABCVV}. 
All other nearest-neighbor  coupling constants equal each other, $D_{i,i+1} =\delta$, $2<i<40$, {  and other coupling constants $D_{ij}$, $|j-i|>1$, are expressed in terms of the nearest-neighbor coupling constants according to their definition given below Eq.(\ref{Hamiltonian})}.  The dimensionless time is $\tau = t \delta$ for this chain.
PST-parameters for the long chains are collected in Table \ref{Table:T2}. { We note that the row $\tau_0$ in this table collects the dimensionless  time instances that are close to $N$, $\tau_0\sim N$. The maxima of  $\lambda^2$ over longer period are not taken into account, although they might exceed those collected in Table \ref{Table:T2}. For instance, for the chain of 40 nodes at $\tau=126.4$ we have $\lambda^2=0.031$. This is a usual compromise problem  between the value of the state-transfer probability  (or the value of $\lambda^2$ in our case) and the time interval required for such transfer.  }
{ All three  values of $\lambda$ found by solving Eq. (\ref{bb3}) at the time instant $\tau_0$ indicated in  Table \ref{Table:T2}  are following {   (for  some values of $N$)}:
\begin{eqnarray}
&&\lambda = 0.265,\; 0.452,\; 0.555, \;\;\;  N=20,\\\nonumber
&&
\lambda=0.136, \;0.268,\; 0.433,\;\;\;N=30,\\\nonumber
&&
\lambda= 0.079, \;0.176,\; 0.204,\;\;\;  N=40,\\\nonumber
&&
\lambda= 0.484, \;0.535,\; 0.741,\;\;\;  N=42_{HFST}.
\end{eqnarray}
Again, only the smallest roots of Eq.(\ref{bb3}) $\lambda_{min}$ are realizable for each $N$ according to Table \ref{Table:T2}.}
\begin{table}
{   \begin{tabular}{|c|cccccccccc|}
\hline
$N$ &10& 20 &25& 30&35 &40&45&50&55&$42_{HFST}$  \cr
$\tau_0$ &14.391&26.506&32.017 &37.393 &42.669&52.846&58.239&63.566&68.839&57.310\cr
$\lambda$ &0.597&$0.265 $ &0.187&$0.136$&0.102 &$0.079$&0.071&0.063&0.056&$0.484 $\cr
$\lambda^2$ &0.356&0.070&0.035 &0.018 &0.010&0.006&0.005&0.004&0.003&0.235\cr
\hline
\end{tabular}}
\caption{Parameters of the $2$-excitation  PST-algorithm  along the long homogeneous  chains of different length  $N$, and along the chain of $N= 42$ nodes with two pairs of coupling constants adjusted for maximizing the parameter $\lambda^2$.
In all chains, the both sender and receiver are  three-qubit  subsystems, $N^{(S)}_2= 3$, and extended receiver is a 5-qubit subsystem ($n^{(ER)}=5$).   We use the unitary transformation $W_{ER}$ of general form preserving the excitation number, Sec.\ref{Section:W}. { We see that  $\lambda^2$ decreases with an increase in the length of the homogeneous chain  and jumps   about  40 times  when passing from the homogeneous 40-node chain  to  the 42-node  chain with two pairs of properly adjusted  end-bonds (weak end-bond control).  The later method was first implemented in the high-fidelity state transfer \cite{GKMT,GMT, FZ_2009,LMDMBA,ABCVV}. This table demonstrates the perspectives of combination of the high-fidelity state transfer (developed in numerous references partially quoted in the Introduction) with the measurement driven PST proposed in our paper.}} 
\label{Table:T2}
\end{table}
{  We see that  $\lambda^2$  decreases  with an increase in the length of the homogeneous chains ($N=20,30,40$) reaching the value $\lambda=0.006$ for $N=40$. 
{   According to Table  \ref{Table:T2}, the dependence of $\lambda^2$ on $N$ in the case of homogeneous chain  can be approximated by the piecewise smooth  function 
\begin{eqnarray}\label{ff}
&&\log (\lambda^2) =\left\{
\begin{array}{ll}
f_1=-0.877N^{0.580} + 2.304, & 10\le N \le 40\cr
f_2=-0.046 N -3.249, & 40\le N \le 55
\end{array}\right.    \;\Rightarrow \\\nonumber
&& \lambda^2 \sim  \left\{\begin{array}{ll}
e^{-0.9 N^{0.6}}, & 10\le N \le 40\cr
e^{-0.05 N} , & 40\le N \le 55
\end{array}
 \right.,
\end{eqnarray}
which is shown in Fig.\ref{Fig:Fig8}a.
\begin{figure*}[!]
\begin{subfigure}{0.45\textwidth}
\includegraphics[scale=0.58]{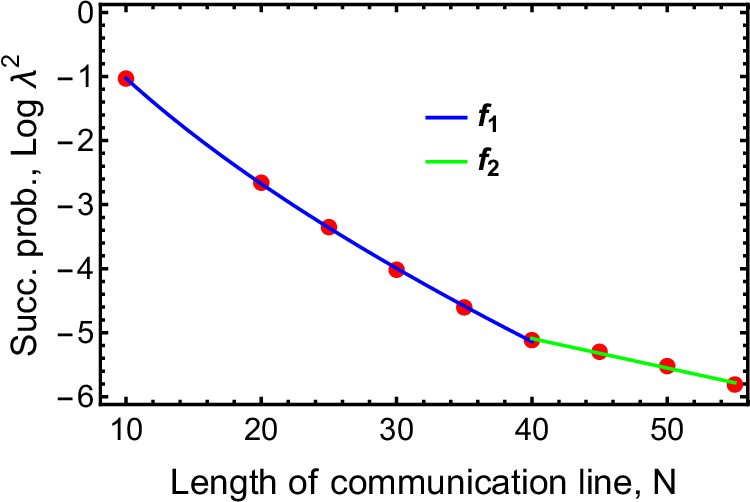}
\caption{}
\end{subfigure}\hfill
\begin{subfigure}{0.45\textwidth}
\includegraphics[scale=0.58]{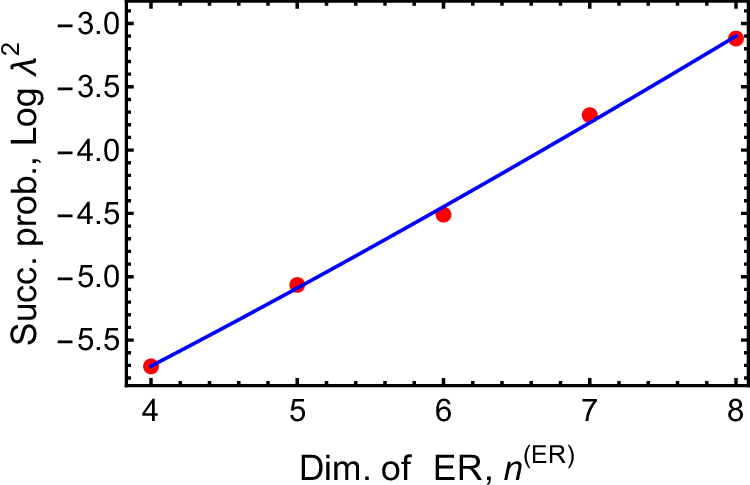}
\caption{}
\end{subfigure}
\caption{{   (a) Dependence of the success probability $\lambda^2$ on the length of the communication line $N$ at fixed dimension of the extended receiver $n^{(ER)}=5$. The smooth approximating functions  $f_1$ and $f_2$ are determined in Eq.(\ref{ff}). (b) Dependence of the success probability $\lambda^2$ on the dimension of the extended receiver $n^{(ER)}$ at fixed  length of the communication line $N=40$. The smooth approximating function  is determined in Eq.(\ref{ff2}).} }  
\label{Fig:Fig8}
\end{figure*}
  {   In principle, the exponential  decrease of $\lambda^2$ with $N$   can be  partially compensated by the exponential increase of  $\lambda^2$ with  the dimensionality of the extended receiver, {   as shown in Table \ref{Table:T3a} for $N=40$ and different sizes of the extended receiver $n^{(ER)}$, see also Fig.\ref{Fig:Fig8}b. 
  \begin{table}
{   \begin{tabular}{|c|ccccc|}
\hline
$n^{(ER)}$ &4 &5&6&7 &8  \cr
$\tau_0$ &57.230&52.846 &55.371 &47.213&49.925\cr
$\lambda$ &0.058 &0.079&$0.105$&0.155&0.210\cr
$\lambda^2$ &0.003&0.006 &0.011 &0.024&0.044\cr
\hline
\end{tabular}}
\caption{{   Increasing $\lambda$ and success probability $\lambda^2$ with $n^{(ER)}$ at fixed length of the communication line $N=40$.  
}} 
\label{Table:T3a}
\end{table} }
Such behavior of  $\lambda^2$ can be approximated by the function 
{   \begin{eqnarray}\label{ff2}
\log (\lambda^2) =
0.377 (n^{(ER)})^{1.203} -7.706 \;\;\Rightarrow \lambda^2 \sim e^{0.4 (n^{(ER)})^{1.2}} .
\end{eqnarray}}
{ However, we suggest an alternative method of  increasing $\lambda^2$  based on the spin-chain prepared for the high-fidelity state transfer. For this purpose we  adjust two pairs of the end-node coupling constants which allow to  significantly increase  $\lambda^2$ till $\lambda^2=0.235$ for $N=42$.  Further increase of $\lambda^2$ via the chain optimization is possible.}

{    { Thus, Tables  \ref{Table:T1}a, \ref{Table:T2} and  \ref{Table:T3a} suggest the two-step optimization of the probabilistic PST before removing the garbage through the ancilla measurement. 

 {\it The first step} is aimed at increasing the success probability of ancilla measurement  and can utilize any known  method  of high-fidelity state transfer (weak end-bond method is utilized in our paper).  A possible resource here is an increase of time interval for state transfer via weakening the end-bonds. Thus, nearly perfect one-qubit state transfer was reached in \cite{GKMT} by increasing twice the distance between the end spins and the body of the chain with appropriate increase in the time interval $\sim N^3$ for the $N$-qubit spin chain with dipole-dipole interactions governed by the $XXZ$-Hamiltonian.   { 
 In Ref. \cite{BZ_2018}, applying the special unitary transformations to the 10-node extended sender and receiver 
 (analogues of the restoring operator  considered in this paper) the excited state was transferred along the $N=776$-node homogeneous spin chain governed by the XX-Hamiltonian  with probability $>0.9$  during time interval $\tau=767.892\sim N$. 
 To provide the larger value of the success probability $\lambda^2$}  in our case of 3-qubit sender/receiver we take all equal coupling constants except the  two pairs: $D_{3,4}= D_{N-3,N-2} =1/8$.  Then, we have for $N=42$:
 \begin{eqnarray}\label{HFST}
 \lambda^2 \approx 0.525 \;\;{\mbox{at}}\;\; \tau_0 = 15720.3. 
 \end{eqnarray}
 
{\it  The second step} is  mainly aimed on the state restoring using the restoring unitary operator applied to the extended receiver according to the strategy of Sec. \ref{Section:rest}.  At that, an increase in dimensionality of the extended receiver also increases the success probability. }

 { In this way, $\lambda^2$ can be  significantly increased, but it can hardly reach 1 except for the case of deterministic PST which is out of consideration here.  Thus, the minimal  number  of copies of the state to be transferred is 2.}  }
}

{\bf Remark.} The exponential decay of $\lambda^2$ with increase in $N$ according to Eq.(\ref{ff})  is caused by the exponential number of states in the garbage $|g_4\rangle$ in Eq.(\ref{Phi42}). The exponential increase of $\lambda^2$ with the increase in $n^{(ER)}$ according to Eq.(\ref{ff2}) is provided by the restoring operator $W_{ER}$ that ``collapses'' the exponential number of states of ER into the state of receiver (coinciding with the initial sender state) according to  restoring system (\ref{Wort32}),  (\ref{Wort312}) with orthogonality  condition (\ref{Wort313}). }

We shall note that, since the success probability of the  ancilla-state measurement  is $\lambda^2$ (see Sec.\ref{Section:OM}),  the probability amplification (Sec.\ref{Section:OM0}) is reasonable in the case $\lambda^2\sim 0.1$ and therefore can be applied to organize the probabilistic  PST along, for instance,  the 42-node chain  with properly adjusted 2 pairs of the end-node coupling constants.

\section{Accuracy  of   algorithm  and some its implementations.}
\label{Section:restoring}
\subsection{Accuracy of state restoring}
\label{Section:W}

{    {We qualitatively discuss the problem of sensitivity of the algorithm to perturbations of the restoring unitary transformation. The restoring parameters $\varphi_0$ can be implemented into the restoring 
 unitary transformation $W_{ER}$ only up to certain accuracy $\varepsilon$, $\varepsilon\ll 1$.}  We can state that  such $\varepsilon$-deviation of the restoring parameters from the predicted values  $\varphi_0$ leads to appearing the ``parasitic'' terms in the restored state. For instance, relation  (\ref{Rnr}) should be replaced with 
 $r_j(t_0,\varphi_0) = \lambda \Big(s_j + \delta(\varepsilon) \sum_{l=0}^{N^{(S)}_k-1} \alpha_{jl} s_l\Big)$, $\delta\ll 1$, and therefore, $|\Psi_{out}\rangle$ in (\ref{PsiOut0})  gets the form 
 $|\Psi_{out}\rangle =\frac{1}{z}\left(\sum_{j=0}^{N^{(S)}_k-1} \Big(s_j |\chi_j\rangle_R +
 \delta(\varepsilon) \sum_{l=0}^{N^{(S)}_k-1} \alpha_{jl} s_l  |\chi_j\rangle_R\Big)\right)$, $z$ is the normalization constant, $z=\sqrt{ \sum_{j=0}^{N^{(S)}_k-1} |s_j + \delta \sum_{l=0}^{N^{(S)}_k-1}  \alpha_{jl} s_l|^2}$, $|z-1| \ll 1$.  Such perturbation terms  reduce the fidelity of the transferred state which now becomes 
 $\frac{1}{z^2}\left|\langle \psi(0)|\Big( |\psi(0)\rangle +  \delta(\varepsilon) \sum_{l,j} \alpha_{jl} s_l  |\chi_j\rangle\Big)\right|^2<1$ ($|\psi(0)\rangle$ is defined in (\ref{Psik})) and tends to 1 with vanishing $\delta$. The success probability of the ancilla-state measurement  now is $(\lambda z)^2$ and does not significantly  differs from $\lambda^2$ because $|(\lambda z)^2-\lambda^2|\ll 1$.
By construction,  the parameter $\delta$ depends only on the local transformation $W_{ER}$ of the extended receiver. It does not depend directly on the parameters of the Hamiltonian. Its relation to $\varepsilon$ can be estimated as $\delta \sim (N^{(ER)}_k) \varepsilon$ in virtue of Eqs.(\ref{RnI}), (\ref{RnT}). Clearly, the increase in $N^{(ER)}_k$ requires the appropriate decrease in $\varepsilon$ to keep $\delta$ the same.
{ We emphasize  that the described accuracy  is a systematic one  with fixed $\alpha_{jk}$, which  can be found knowing the accuracy of realization of basis unitary transformations and exact values of restoring parameters $\varphi_0$.  In this case, the parameters $\alpha_{jk}$ can be included into the state-restoring protocol. }
 It is important that  the above ``parasitic'' terms are generated  locally by the restoring transformation  of  the extended receiver while other nodes of the communication line do not effect on  these terms. This is the principal advantage of our {   probabilistic} PST-algorithm over the {   deterministic} PST-algorithms based on the completely engineered chains quoted in the Introduction, see, in particular,  Refs.\cite{CDEL,KS}. Therefore, we can reduce or even vanish these terms via a local tool  increasing the accuracy of the restoring unitary transformation. This is  a reasonable effort because the {   required} dimensionality of the restoring  transformation (or  the dimensionality of the extended  receiver) is defined only by the dimensionality of the transferred state (initial state of the sender) and does not depend on the length of the communication line $N$.  {   Consequently,  we can select the $\hat n^{(ER)}$-qubit subsystem at the receiver side   (maximal extended receiver)  and prepare the basis set of the  high-accuracy unitary transformations acting  in the state-space of this subsystem. These transformations can be used to construct the restoring operator $W_{ER}$ 
 for restoring the state transferred along the spin chain of any length $N$ provided that the required dimensionality of the extended receiver $n^{(ER)}$ does not exceed  $\hat n^{(ER)}$.}}
 
 \subsection{Effect of quantum noise}
{    The noise of quantum gates leads to another type of error. {   Although the formulae for the perturbed transferred state and for the state-transfer fidelity are similar to that in Sec.\ref{Section:W}, now the perturbations are generically spread over the whole  $2^{n^{(ER)}}$-dimensional state space of the extended receiver}, the parameters $\alpha_{jk}$ are defined by the quantum noise and are  not  fixed in this case, therefore they can not be included into the state restoring protocol.   The quantum gate noise  is the usual problem of contemporary quantum computations.}}
 
 {   Regarding the accuracy of realization of the state-encoding (decoding) subroutines, these operations use only $CNOTS$ with multiqubit control and therefore the accuracy of their realization is completely defined by the quantum noise of the two-qubit operations. The effect of quantum noise  can be partially overcome  using the post-treatment at the receiver side, proposed, for instance, in  \cite{ZRPL,BP}.
 
 \subsection{State transfer and restoring with unknown Hamiltonian}
Another aspect which has to be discussed is  the   implementation  of the algorithm to the spin-system governed by the unknown excitation-preserving Hamiltonian. We can refer the excitation-preserving Hamiltonians  with unknown coupling constants to such type of Hamiltonians.  
 As the  preliminary action, one has to perform the set of  test experiments by transferring the  basis states of the sender state-space and performing the state-tomography of the extended receiver state   at some time instant, thus  revealing the elements of the matrix $\hat V$ according to Eqs.(\ref{RnI}),  (\ref{RnT}) with disregarded $W_{ER}$:  $W_{ER}=I$ (identity matrix). Knowing $\hat V$,  we define $\lambda^2$  at that time instant solving the polynomial equation (\ref{lamDp}).  Scanning the time interval we can find the  time instant maximizing  $\lambda^2$ and refer to this time instant as the time  instant for state registration $t_0$. This is a cumbersome procedure but it must be performed only once for the particular   unknown Hamiltonian.  After that, the restoring transformation $W_{ER}$ can be constructed solving   the restoring system  (\ref{Wort32}),  (\ref{Wort312}) with orthogonality  condition (\ref{Wort313}).}


{   
\subsection{Applicability to superconducting quantum platform}

The multiqubit  state transfer along the spin chain can be  a valuable operation in information exchange inside a given quantum platform.  
In this section we discuss some aspects of realizability of the probabilistic PST  using the flux-qubit dynamics on a superconducting quantum platform.  It was shown in \cite{MREKTA} that the flux-qubit evolution can be described by the approximate $XX$-Hamiltonian 
\begin{eqnarray}\label{HSP}
H=\frac{1}{2}\sum_{j=1}^N \Delta_j \sigma^{(z)}_j + \frac{1}{2} \sum_{i<j; i,j=1}^N g_{ij} (\sigma^{(x)}_i \sigma^{(x)}_j + \sigma^{(y)}_i \sigma^{(y)}_j),
\end{eqnarray}
 where $\sigma^{(\alpha)}_j$, $\alpha=x,y,z$, are the Pauli matrices, $\Delta_j$ are the energy gaps between the  states $|0\rangle$ and $|1\rangle$ of the $j$th qubit, $g_{ij}$ are the coupling constants, which  can be tuned    with a special microwave-driving tool \cite{NHYNLT}.  If $\Delta_j=\Delta$, then the first term in (\ref{HSP}) disappears in the rotating frame of reference.  In Eq.(\ref{HSP}),  the coupling constants $g_{ij}\equiv D_{ij}$.  To excite the particular qubit, we first  set all the qubits to the ground state by cooling  and then apply  a resonant microwave $\pi$-pulse
 using the   microwave line connected to the appropriate  qubit \cite{CW,CNHM}. Formally, applying  the $\pi$-pulse means  applying the $\sigma^{(x)}$-operator to the qubit. We also mention that the flux-qubit  dynamics in  the superconducting chain for  the Dicke model was considered in \cite{RZPL}.
 
 Realizability of the state-transfer on the superconducting platform is defined by the comparison of the state-registration time instant $t_0$ with the decay time $T_1$ in superconducting platform: $t_0\ll T_1$. We refer to the data on quantum simulator  in Ref.\cite{MREKTA}, Table A.1., where $T_1\approx 10^{-5} s$, the coupling constant ($D_{12} $ in our case)  $\sim 10^9 Hz$. Since $\tau_0 = t_0 D_{12} $, then $ \tau_0 \ll T_1 D_{12} \sim 10 000$. Comparison of this estimation with the time intervals $\tau_0$ in Tables \ref{Table:T1} --  \ref{Table:T3a}   predicts the realizability of our algorithm for the chains of about hundreds qubits.  After the successful measurement of the ancilla state, the transferred state $|\Psi_{out}\rangle$ in Eq.(\ref{Phi32}) (in the case of the $k$-excitation state transfer, see also Fig.\ref{Fig:1ex}) or in Eq.(\ref{Psiout0}) (in the case of  an  arbitrary state transfer, see also Fig.\ref{Fig:arb}) can be used in further computation. { An addition restriction on the maximal chain length is the success probability which is rapidly vanishing with an increase in  $N$ in homogeneous chains. For instance, assuming validity of (\ref{ff}) for larger $N$ we obtain $\lambda^2 \sim 7\times 10^{-3}$  for $N=100$.  This obstacle can be overcame using the two-step optimization of the probabilistic PST proposed in Sec. \ref{Section:num}, thus utilizing  the chains adjusted for  large success probability like the chains with weak end-bonds (see the 42-node chain in Table \ref{Table:T2} and the values in Eq.(\ref{HFST})). The possibility to control the individual coupling constants on the superconducting quantum  platform  via  the  microwave line connected to the appropriate  qubit  can be effective tool for organizing large $\lambda^2$ at the first step of the optimization algorithm}
}

\section{Conclusions}
\label{Section:conclusions}
We present the algorithm  for the PST which combines the state restoring protocol with the ancilla measurement. The simplest variant of the algorithm for the perfect transfer of an arbitrary state   includes all spins of the chain into the controlled operator labeling the garbage, which is not convenient for applications. However, this disadvantage disappears in the PST of  so-called $k$-excitation states, which are  superpositions of states with the same excitation number $k$.  Therefore, the detailed algorithm for the  PST of the $k$-excitation states is  presented in Sec.\ref{Section:FixEx}.  
The final step in  the PST-algorithm is the  measurement of the state of the ancilla $B$ with the success probability $\lambda^2$. If  $\lambda^2$ is large enough, the amplification probability might be  effective. 

We also show that the arbitrary state of the subsystem $S^{(0)}$ can be perfectly transferred {after been encoded} into the $k$-excitation state of $S$ and then, after restoring at the receiver$R$, been decoded  into the state of $R^{(0)}$ that coincides with the initial state of $S^{(0)}$. It is important that such encoding requires minor increase in the dimensionality of $S$  ($R$) in comparison with that of  $S^{(0)}$ ($R^{(0)}$). 

The restoring protocol for the $k$-excitation pure  state is presented in details. It uses the unitary transformation preserving the excitation number. 
However, the transformation that does not preserve the excitation number can also be effective but requires certain modification of the algorithm and is given in Appendix, Sec. \ref{Section:A2}}. { It is remarkable that  the maximized  $\lambda$ is the same for both cases and   can be predicted theoretically.}

{ We note that  the $k$-excitation state transfer  requires certain post-processing over qubits of the extended receiver at the time instant $t_0$ including state restoring and ancilla measurement, while evolution of all other qubits of the communication line is not subjected to any additional action. Therefore, the fidelity of the transfer state is completely defined by the quality of operations over  the confined number of qubits of the extended receiver which simplifies the control.  In addition, dealing with  an arbitrary  state transfer, one has to evaluate the  encoding  subroutine at the sender side (the preliminary actions before  initiating the state transfer)  and decoding subroutine at the receiver side (the post-action after state restoring). Quality of the operators used in encoding-decoding processes are also responsible for  the fidelity of the transferred state.   Again however,  in both subroutines,  only operations with confined number of qubits is involved and this number does not depend on the length of the considered communication line. We also note that the subsystems $S^{(0)}$ and $R^{(0)}$ included into the communication line and containing the  state to be transferred ($S^{(0)}$) and the transferred state ($R^{(0)}$) can be moved out of the communication line and considered as the additional auxiliary subsystems not involved into the Hamiltonian evolution. The same holds for the nodes of the extended receiver $ER$ which can be also moved out of the communication line. In this way,  we completely remove the state-destroying by the qubits of transmission line $TL$  imposing the responsibility for  the fidelity of state transfer  on the accuracy of the operations over the subsystems  $S$, $R$ and  extra subsystems $S^{(0)}$, $R^{(0)}$, $ER$. However, we do not include such technical details in the manuscript   which presents the basic mathematical model as the foundation  for the proposed approach to the  problem of multiqubit pure PST.}

{ 
Now we collect the basic features of the protocol in view of obtained results. 
\begin{enumerate}
\item
Since the effective PST-algorithm is closely related to the  large enough absolute value of the $\lambda$-parameter, we are interested in the tool allowing to increase $\lambda$. Along with  increasing in the dimension of the extended receiver, this  goal can be reached using inhomogeneous chains, like the chain of 42-nodes considered in Sec.\ref{Section:num}, see Table \ref{Table:T2}, { or, alternatively, any algorithm allowing the  high-fidelity state transfer. Thus, only the restricted number of spins are involved in the control process. Those are the spins of the extended receiver and several spins (two in our example) spins at the sender side forming the  properly adjusted  weak bonds}. {    Some disadvantage of such tool of HFST is that weakening the end-bonds causes the increase in the optimal time instant for state registration, which can be significant in certain cases, see the parameters in (\ref{HFST}). } 
\item
The probability of  PST $\lambda^2$ estimates the number { $O(\lambda^{-2})$ }of runs needed to succeed in this process. However, the fidelity of the transferred state in the successful case is always one, i.e., the receiver's state coincides with the initial sender's state. {    The weakness here is that $O(\lambda^{-2})$-running requires preparing $O(\lambda^{-2})$-copies which is possible only for  a known state.  
{ Instead of multiple running of the single algorithm, we can use the single run of  the $O(\lambda^{-2})$-channel  algorithm, where 
 each channel runs the same algorithm including the state transfer and restoring as a subroutine  with subsequent postselection after ancilla measurement in each channel. This does not remove the problem of preparing the 
 $O(\lambda^{-2})$ copies, but allows to shorten the algorithm evaluation-time. However,
 there is a type of problems   when not all states of the sender are necessary to be registered in the receiver. A simple example is the quantum random-number generator \cite{MMP}, where the output is a random state.
 If the output register of such generator is the sender $S$ of our communication line, then the probabilistic PST means that  any particular 
 generated  random state will be transferred to the receiver $R$ with the probability $O(\lambda^2)$, i.e., not each  random state reaches the receiver $R$. Creating $O(\lambda^{-2})$  copies of each  random state to be transferred is not necessary in this case, just the number of transferred random states  will be less then the number of generated ones.} }
\item
There is no principal requirement to dimensionality of the transferred state. An increase in the state-dimensionality requires larger extended receiver and reconstructing the restoring unitary transformation. {   However, generically an increase in  dimensionality of  the transferred state causes the increase in the excitation number $k$ used in the communication line and  in the dimension of the whole state-space. As a consequence, the  success probability $\lambda^2$  generically decreases. }
\item
We shall emphasize that 
our algorithm is not sensitive to perturbations of the Hamiltonian provided that these perturbations preserve the excitation number in the system. The evolution under such perturbed Hamiltonian supplemented with the appropriate state restoring unitary operator can effect only the   value of the  parameter $\lambda$ which defines the probability of PST and  disappears after ancilla-state  measurement. 
\item
{    The probabilistic  PST-algorithm can be prepared for the communication line governed by the unknown excitation-preserving Hamiltonian.}
\item
We emphasize that, if the conditions of  Proposition 2 are satisfied,  the problem of maximizing the probability $\lambda^2$ via parameters of the restoring unitary transformation is reduced to solving the additional polynomial equation imposed on $\lambda^2$, thus avoiding the  time-consuming   numerical maximization. 
\end{enumerate}
}

The circuits for all proposed algorithms are present and discussed.

{\bf  Acknowledgments.} 
\newline
{    The authors thank the anonymous referees for useful comments.} 
This project is supported partially by the
National Natural Science Foundation of China (Grants No. 12031004, No.
12271474). The work is partially funded as a state task of Russian Fundamental Investigations (State Registration No. 124013000760-0).


{  
\section{Appendix. Operator $W_{ER}$ not preserving excitation number}
\label{Section:Appendix}
\label{Section:A2}
It is interesting that  restoring  the $k$-excitation state used  in Sec.\ref{Section:FixEx}  admits the  operator $W_{ER}$   which does not preserve the excitation number. Realization   of such operator is proposed, for instance, in  Ref.\cite{WSW}. 
The circuit for state transfer must be slightly modified and the additional one-qubit ancilla $D$ must be included, see Fig.\ref{Fig:NCc}.
\begin{figure*}[!]
\centering
\includegraphics[scale=0.5]{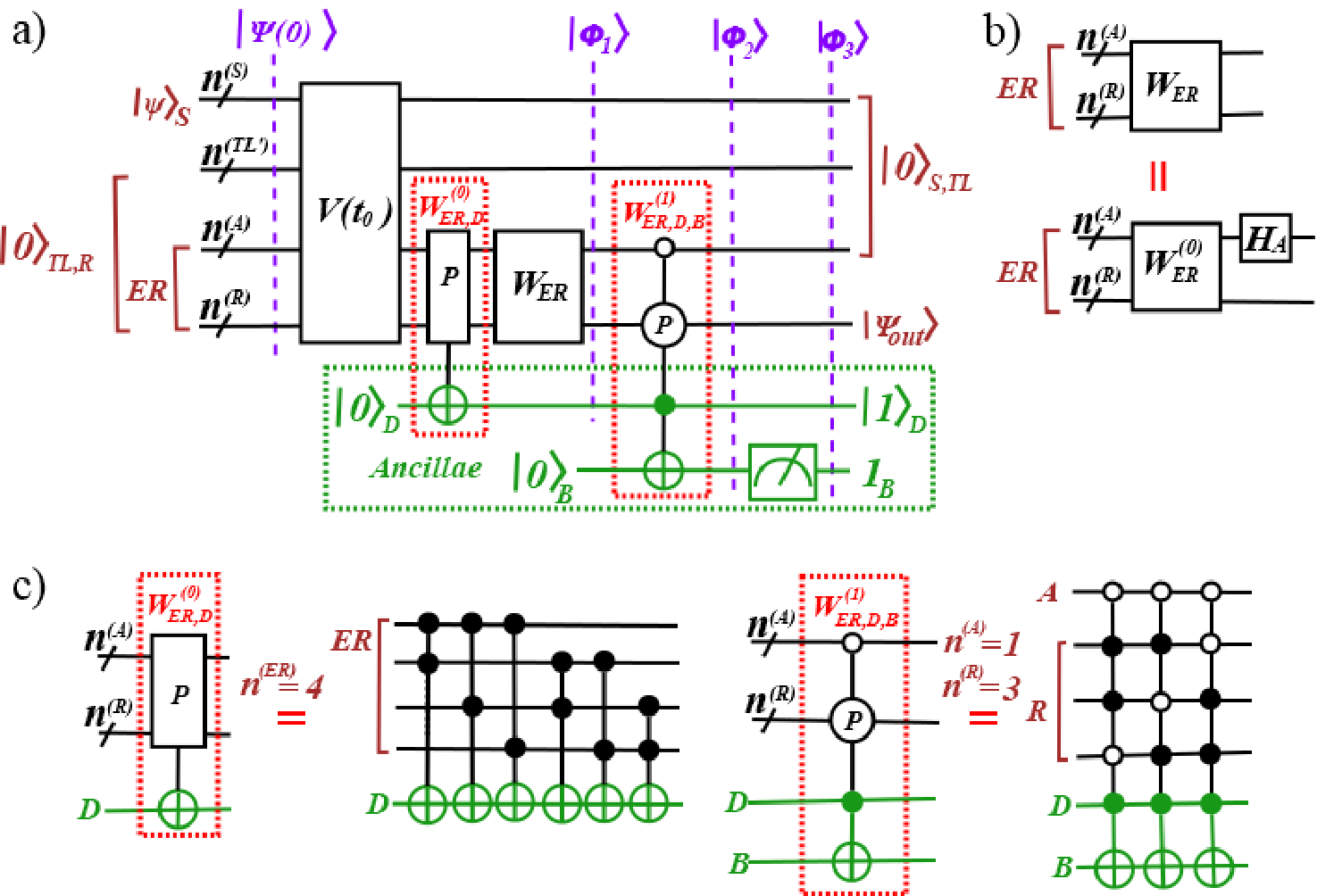}
\caption{(a)The circuit for the perfect $k$-excitation state transfer with the operator $W_{ER}$ not preserving the excitation number. { Here $|\Psi(0)\rangle$ is the initial state, the intermediate states $
|\Phi_j\rangle$, $j=1,2,3$, are defined, respectively, in  Eqs.(\ref{D}), (\ref{Phi2})  with replacement (\ref{W1ARB}) and (\ref{Phi32}) with $|\Psi_{out}\rangle$ defined  in Eq.(\ref{PsiOut0}).}  (b) The structure of the operator $W_{ER}$. (c) The operators $W^{(0)}_{ER,D}$ and  $W^{(1)}_{ER,D,B}$ for the two-excitation case, { 3-qubit receiver ($n^{(R)}=3$) and 1-qubit ancilla $A$ ($n^{(A)}=1$), i.e., the extended receiver $ER$ includes 4 qubits.} }  
\label{Fig:NCc}
\end{figure*}
In this case,
before applying  $W_{ER}$, we apply the operator 
$W_{ER,D}^{(0)}$,
\begin{eqnarray}&&
W_{ER,D}^{(0)}= P_{ER}\otimes \sigma^{(x)}_D + (I_{ER}-P_{ER})\otimes I_D,\\\label{P}
&&
P_{ER}=\sum_{j=0}^{N^{(ER)}_k-1}
|\chi_j\rangle_{ER_{j}} \, {_{ER_{j}}\langle\chi_j|},
\end{eqnarray}
which labels the states needed in the subsequent transformations.  This labeling is necessary to guarantee that only the $k$-excitation states of the $ER$  at the input of $W_{ER}$ will be further transformed. Recall that $|\chi_j\rangle_{ER_{j}}$ is the product of $k$ excited one-qubit states included in the $k$-excitation state $|\chi_j\rangle_{ER}$. 
{ The depth of $W_{ER,D}^{(0)}$ is $O(k N^{(ER)}_k)$.}
Taking into account the operator $W_{ER,D}^{(0)}$, the  formula for $|\Phi_1(t_0,\varphi)\rangle\equiv |\Psi(t_0,\varphi)\rangle$, defined in Eq.(\ref{Phi1star2}), must be modified as follows:
\begin{eqnarray}\label{D}
 |\Phi_1(t_0,\varphi)\rangle = W_{ER}(\varphi)W_{ER,D}^{(0)} \left(e^{-i H t_0}|\Psi(0)\rangle\right)\otimes |0\rangle_D   = {|0\rangle_{S,TL}|\psi(t_0,\varphi)\rangle _R  |1\rangle_D  +|g_1\rangle ,}
\end{eqnarray}
where 
\begin{eqnarray}\label{WH}
W_{ER}= H_A W_{ER}^{(0)} ,
\end{eqnarray}
and  $W_{ER}^{(0)}$ is the state-restoring unitary transformation,  $H_A$ is the set of the Hadamard operators applied to the qubits of  $A$, see Fig.\ref{Fig:NCc}b. 
We emphasize, that the qubit $D$ in Eq.(\ref{D}) is not subjected to the evolution governed by the Hamiltonian $H$. It is included into the algorithm at the time instant of state registration  $t_0$.
Next, the projector $P_{R}$ in Eq.(\ref{PRj}) and the controlled operator $W^{(1)}_{RB}$ in Eq.(\ref{W1RB1}) must be modified as follows:
\begin{eqnarray}\label{PARj}
&&
P_{R} \to  P_{A,R,D}=
|0\rangle_A\,{_A\langle 0|}\sum_{j=0}^{N^{(S)}_k-1} |\chi_j \rangle_{R}  \,{_{R}\langle \chi_j|}  \otimes|1\rangle_{D}\,{_{D}\langle 1|}  ,
\\\label{W1ARB}
&&
{   W^{(1)}_{RB} \to 
W^{(1)}_{ER,D,B}=P_{A,R,D}\otimes \sigma^{(x)}_{B} + (I_{A,R,D}- P_{A,R,D}) \otimes I_{B}.}
\end{eqnarray}
{ The depth of $W^{(1)}_{ER,D,B}$ is $O( n^{(ER)}N^{(S)}_k )$ instead of the depth $O(k N^{(S)}_k)$ of the operator $W^{(1)}_{RB}$ defined in Eq.(\ref{W1RB1}).}
We emphasize the difference in the states used in projectors (\ref{PRj}) and  (\ref{PARj}): 
the tensor product of $k$ one-qubit excited state  $|\chi_j\rangle_{R_j}$ is used  in Eq.(\ref{PRj}) and the $n^{(S)}$-qubit $k$-excitation state $|\chi_j\rangle_{R}$  is used  in Eq.(\ref{PARj})   for each $j$.  This difference is caused by the fact that   the operator $W_{ER}$ in this section does not preserve the excitation number, so that the excitation number may be bigger then $k$.
 The pointed difference is also reflected in Figs.\ref{Fig:1ex}b,c and \ref{Fig:NCc}c. 
 
 Thus, using  the operator  $W^{(1)}_{ER,D,B}$ instead of $W^{(1)}_{RB}$ in Eq.(\ref{W1}) we write Eq.(\ref{Phi2}) as follows:
 \begin{eqnarray}\label{Phi22}
|\Phi_2\rangle =  W^{(1)}_{ER,D,B}|\Phi_1\rangle |0\rangle_B = |0\rangle_{S,TL} |\psi(t_0,\varphi_0)\rangle_R |1\rangle_{D} |1\rangle_{B} + |g_1\rangle |0\rangle_B,
\end{eqnarray}
where $|\psi(t_0,\varphi_0)\rangle_R$ is defined in Eq.(\ref{restore2}).
 {The depth of the operator $W^{(1)}_{ER,D,B}W_{ER}W^{(0)}_{ER,D}  $  is 
\begin{eqnarray}\label{OO}
O(n^{(ER)} N^{(S)}_k +{\mbox{depth}} (W_{ER})+ k N^{(ER)}_k ).
\end{eqnarray}
Measuring the state of the ancilla $B$ with the desired output $|1\rangle_B$ (success probability is $\lambda^2$) we end  up with the state $|\Phi_3\rangle$ defined in Eqs. (\ref{Phi32}), (\ref{PsiOut0}) {up to ignoring the state $|1\rangle_D$}. The number of runs of the algorithm needed to get the successful measurement can be estimated as $O(\lambda^{-2})$.
Then, instead of Eq.(\ref{dQ}),  we have the following expression for the  time required for the PST:
  \begin{eqnarray}\label{qQ120}
&&T^{(PST)}=\\\nonumber
&&O\left(\Big(   N^{(S^{(0)})} k n^{(S^{(0)})}t^{(op)} + t_0 +
( N^{(S)}_k n^{(ER)}+{\mbox{depth}} (W_{ER}) +   k N^{(ER)}_k   )t^{(op)} \Big)  \lambda^{-2}\right).
\end{eqnarray}}
Since the operator $W_{ER}$ in the form of Eq.(\ref{WH}) doesn't preserve the excitation number in the system,  the higher order excitations will be generated   after applying  such operator. However, the controlled operator  $W^{(1)}_{ER,D,B}$, defined in Eq.(\ref{W1ARB}), puts these higher-excitation terms into the garbage  $|g_1\rangle$.  

\subsection{General form of  operator  $W_{ER}$ not preserving excitation number}
\label{Section:W2}
  Due to the fact that we deal with the $k$-excitation state transfer and consider only those terms in the  superposition states $|\Phi_3\rangle$, given in  Eq. (\ref{Phi32}), which collect all excitations in the  receiver $R$, we can select the  block in the product $W V$ (see Eq.\ref{Psi0}) that transfers $|\psi(0)\rangle_S$, defined in Eq.(\ref{Psik}), from the initial  state $|\Psi(0)\rangle$, defined  in Eq.(\ref{PsiIn}),  to the receiver state $|\psi(t_0,\varphi_0)\rangle_R$ in  {the final state} $|\Phi_3\rangle$, given  in Eq.(\ref{Phi32}). 
  
In this section, similar to Sec.\ref{Section:W}, we use the  {  multiindexes} with subscripts indicating the appropriate subsystem.
Eq. (\ref{RnI}) for $r_{j_R}$ holds for this case as well.
Since we deal with the $k$-excitation initial state, only $k$-excitation block of $V$ is used.  In the receiver state, we also  need only the $k$-excitation states. These two conditions propose the   expression (\ref{RnT}) for  $T^{(k)}_{j_Rm_S}$.
However, now  $W_{ER}^{(k)} (W_{ER}^{(k)})^\dagger \neq I_{ER}$ because $W_{ER}$ does not have the block-diagonal form (\ref{WW20}), unlike the excitation preserving  case.
Introducing matrices $\hat W_{ER}$ and $\hat V$ with the elements
$(\hat W_{ER})_{j_R;n_{ER}}  =(W_{ER}^{(k)})_{0_Aj_R;n_{ER}}$ and $\hat V_{n_{ER};m_S} = V^{(k)}_{0_S0_{TL'} n_{ER}; m_S0_TL0_R}$, we can write 
$\vec r = \hat W_{ER} \hat V \vec s$, where $\vec r$ and $\vec s$ are  the column-vectors with entries, respectively, $r_{j_{R}}$ and $s_{m_S}$. In addition, the unitarity condition holds:
\begin{eqnarray}\label{AWort}
 W_{ER}  W_{ER}^\dagger = I_{ER},
\end{eqnarray}
where $I_{ER}$ is the identity matrix acting in the whole state space of $ER$. 
The restoring constraints (\ref{Wort31}) and (\ref{Wort32}) remain the same.
The optimization parameters are the elements of $\hat W_{ER}$, which must satisfy constraints (\ref{Wort31}) and (\ref{Wort32}) supplemented with the unitarity condition (\ref{AWort}).
{Thus the list of effective  parameters $\varphi$ in $W_{ER}$  reads
\begin{eqnarray}\label{varphi2}
\varphi =\{ {\mbox{Re}}( W_{ER})_{jn}, {\mbox{Im}}(W_{ER})_{jn}:  j,n=0,\dots, 2^{n^{(ER)}}-1\}.
\end{eqnarray}
{We use the parameters of the unitary transformation $W_{ER}$ not only to restore the transferred state, but also to  increase $\lambda$.  For this purpose, similar to the excitation preserving case, we  reduce the number of nonzero probability amplitudes in the garbage $|g_2\rangle$ in Eq.(\ref{Phi22}). Since the unitary transformation $W_{ER}$ generates all possible excitations, the  subscript $i_{ER}$  in   $(W_{ER})_{i_{ER};j_{ER}}$ can run the basis vectors with all possible excitation numbers.   Thus, we introduce the matrix $\overline W$ with the elements {$(W_{ER})_{i_{ER}; j_{ER}}$, $i_{ER}=0,\dots,2^{n^{(ER)}}-1$, $j_{ER}=0,\dots, N^{(ER)}_k-1$ ($j_{ER}$ runs the basis vectors of the $k$-excitation 
state subspace)} and    extend   Eq.(\ref{Wort31}) as follows: 
\begin{eqnarray}\label{AWort312}
  &&( \overline W\hat V)_{i j_R}=  0\;\; \Leftrightarrow \;\; a_i b_{j_R} =0, \;\; i \neq  j_R, \\\nonumber
  && i = 0,\dots, 2^{n^{(ER)} }- N^{(S)}_k ,\;\; j_R=0,\dots, N^{(ER)}_k-1,
\end{eqnarray}
which generalizes constraints (\ref{ab2}) written for the excitation-preserving case. Obviously,   constraints  (\ref{Wort31}) are included into Eq.(\ref{AWort312}).
The upper limit for $i$ in Eq.(\ref{AWort312}) is defined by the requirement that 
the number of constraints, which can be added to  Eq.(\ref{Wort31}), equals  $N^{(ad)} = 2^{n^{(ER)} }- N^{(ER;min)}_k= 2^{n^{(ER)} } - 2 N^{(S)}_k+1$, so that 
the upper limit for $i$ equals $N^{(S)}_k-1 +N^{(ad)} =2^{n^{(ER)} }- N^{(S)}_k$ .  We emphasize, that the basis of $2^{n^{(ER)}}$ states is ordered by the excitation number as follows:
\begin{eqnarray}\label{basisk}
&&|0\rangle,\;\\\nonumber
&& |1_j\rangle, \;\; j=1,\dots, n^{(ER)}, \\\nonumber
&& |1_j1_l\rangle, \;\; j=1,\dots, n^{(ER)}-1,\; l=j+1,\dots,  n^{(ER)},\\\nonumber 
&& \dots, \\\nonumber
&&  |1_1\dots 1_{n^{(ER)}}\rangle.
\end{eqnarray}
}
In this case Proposition 2 holds, so that the value $\lambda$  is defined by the polynomial equation  (\ref{lamDp}) as well and does not depend on the particular restoring unitary transformation. { Consequently, the success probability $\lambda^2$ does not depend on whether $W_{ER}$ preserve the excitation number in the system or  does not preserve.} 

\subsection{Particular realization of operator $W^{(0)}_{ER}$}
\label{Section:AWpart}
We use the unitary transformation $W_{ER}$ in the  form of Eq.(\ref{WH}), see Fig.\ref{Fig:NCc}b, and present a version of the operator $W_{ER}^{(0)}$ which is  a modified version of the unitary transformation proposed in  Ref.\cite{WSW}.  
Namely,
we represent $W_{ER}^{(0)} $ as follows:
\begin{eqnarray}\label{WH2}
W_{ER}^{(0)} =  V^{(Q)}\dots V^{(1)},
\end{eqnarray}
where
\begin{eqnarray}\label{Vk}
V^{(k)}=\prod_{i=1}^{n^{(ER)}-1} \prod_{j=i+1}^{n^{(ER)}}      C_{ij}  \prod_{l=1}^{n^{(ER)}} U^{(k)}_{l},
\end{eqnarray}
\begin{eqnarray}\label{Cjk}
C_{ij}=|1\rangle_i\, {_i \langle 1|} \otimes \sigma^{(x)}_j + |0\rangle_i\, {_i \langle 0|} \otimes I_j .
\end{eqnarray}  
Here the subscripts $i$, $j$ and $l$ mean the appropriate  qubits of $ER$.
The circuit  for the discussed operator $W_{ER}^{(0)}$ acting on  the 4-qubit extended receiver, $n^{(ER)}=4$, is shown in Fig.\ref{Fig:W}.
\begin{figure*}[!]
\centering
\includegraphics[scale=0.5]{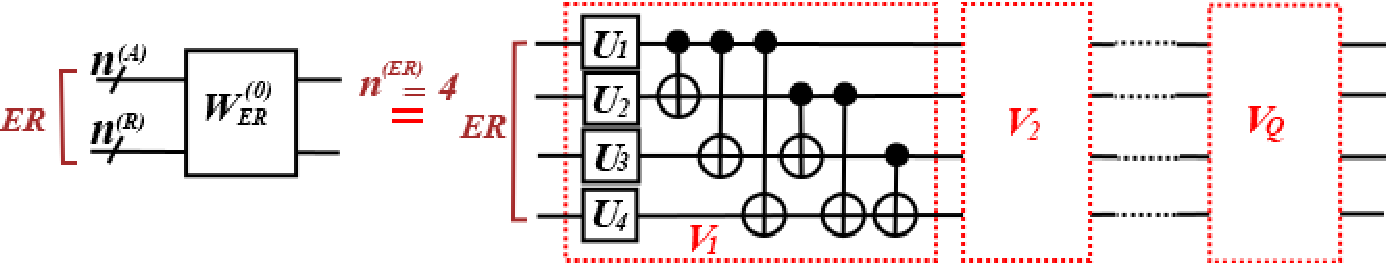}
\caption{The unitary transformation $W^{(0)}_{ER}$, { defined in Eqs.(\ref{WH2}) - (\ref{Ukj}),} not preserving the excitation number.  Example is given for $n^{(ER)}=4$. }  
\label{Fig:W}
\end{figure*}

{ The depth of $W_{ER}$ can be estimated by the depth of  $W_{ER}^{(0)}$ and equals $O(Q (n^{(ER)})^2)$}.
For the unitary operators $U^{(k)}_j$ in Eq.(\ref{Vk}), we use the following representation in terms of one-qubit rotations:
\begin{eqnarray}\label{Ukj}
U^{(k)}_j &=& { e^{i \mu_{kj}}} R_{zj}(\alpha_{kj}) R_{yj}(\beta_{kj}) R_{zj}(\gamma_{kj}),\\\nonumber
&&j=1,\dots, n^{(ER)},\;\; k=1,\dots, Q.\end{eqnarray}
Thus, the list of parameters in $W_{ER}$ is 
\begin{eqnarray}\label{Avarphi}
\varphi= \{ \alpha_{kj}, \beta_{kj},\gamma_{kj},\mu_{kj}: j=1,\dots,  n^{(ER)}, \;\; k=1,\dots,Q\}.
\end{eqnarray}
Eq.(\ref{qQ120}) for the running time of the  PST-algorithm  now reads:
  \begin{eqnarray}\label{qQ12}
T^{(PST)}&=&\\\nonumber
&&O\left(\Big(   N^{(S^{(0)})} k n^{(S^{(0)})}t^{(op)} + t_0 +
( N^{(S)}_k n^{(ER)}+Q(n^{(ER)})^2 +   k N^{(ER)}_k   )t^{(op)} \Big)  \lambda^{-2}\right).
\end{eqnarray}
It is important that Eqs.(\ref{bb1}), (\ref{bb3}), (\ref{lamDp}) hold in this case as well. 
 
{\subsection{Numerical restoring of  3-qubit 2-excitation state}
Similar to Sec.\ref{Section:num}, we consider the 2-excitation 3-qubit state restoring using 10-node spin-1/2 chain, $N^{(ER)}=4,5,6$.
The restoring system is  represented by  Eqs.(\ref{AWort}),  (\ref{Wort32}),  (\ref{AWort312}). 
For calculation the exact value of $\lambda$, Eq.(\ref{bb3}) is applicable in this example yielding values 
$\lambda$ and $\tau_0$ in Eq.(\ref{lll}) and in Table \ref{Table:T1}.  Therefore, the case of generic matrix $W_{ER}$ not preserving the excitation number yields the same result as the case of excitation preserving $W_{ER}$.  Now  we consider 
the  particular realization of $W_{ER}$ proposed in Sec.\ref{Section:AWpart}.  In this case,   the number of free parameters $\varphi$ in list (\ref{Avarphi}) is not enough to satisfy all constraints (\ref{AWort312}).
Therefore, instead of Eq.(\ref{AWort312}), we  combine constraints (\ref{Wort31}) with  some other constraints on the two-excitation subspace  written as
\begin{eqnarray}\label{AWort3132}
  &&a_{i_R}b_{j_R} =0, \\\nonumber
  && j_R=0,\dots, N^{(S)}_2-1,\;\; i_R= N^{(S)}_2,\dots,  N^{(S)}_2+\tilde N^{(ad)}_2-1 ,\;\; i\neq j_R.
\end{eqnarray}
 {Since the condition of  Proposition 2 is not satisfied, $\lambda$ depends on the elements of $W_{ER}$. The maximization of $\lambda$ is performed over 1000 solution of the restoring system  (\ref{AWort}), (\ref{Wort31}),  (\ref{Wort32}), (\ref{AWort3132}),  each  solution  yields different $\lambda$. The results are collected in Table \ref{Table:T3}.    
The parameter $\tilde N^{(ad)}_2$ is indicated  for each $N^{(ER)}$. Here $\lambda$  decreases with an increase in $N^{(ER)}$ which  happens because the fraction of the vanishing probability amplitudes in the garbage (defined by the parameter $\tilde N^{(ad)}_2$) decreases with an increase in $N^{(ER)}$. This is justified by comparison of $N^{(ad)}$ with $\tilde N^{(ad)}_2$ in Table \ref{Table:T3}. However, unlike the excitation preserving $W_{ER}$, see Table \ref{Table:T1}b,  $\lambda$ increases with $Q$ only at  $n^{(ER)}=4,5$, while $\lambda$ at  $n^{(ER)}=6$ is almost the same for $Q=2,3$. Such behavior is explained by large number of free parameters  in $W_{ER}$ which makes an obstacle for finding the global maximum.  { Comparison of $\lambda^2$  in Tables \ref{Table:T1} and \ref{Table:T3} shows the advantage of the excitation preserving unitary transformation of the extended receiver $W_{ER}$ over the transformation $W_{ER}$ not preserving the excitation number. }}
\begin{table}
\begin{tabular}{|c|cccccc|}
\hline
$n^{(ER)}$ & 4 & 4 & 5&5&6&6 \cr
$Q$            &2  &3  &2&3&2&3\cr
$N^{(ER)}_2$ & 6&6&10&10 &15&15 \cr
$N^{(ad)}$ & 11&11& 27&27 &59&59 \cr
$\tilde N^{(ad)}_2$ & 0&2& 1&3 &2&5 \cr
$\tau_0$ &12.493&12.493&14.391&14.391&14.132&14.132\cr
$\lambda$ & $0.257 $&$0.281$ &$0.224 $&$0.227$ &$0.182$&$0.181 $   \cr
$\lambda^2$ &0.066                   &0.079 &0.050&0.152  &0.033&0.033 \cr
\hline
\end{tabular}
\caption{Parameters for the $2$-excitation 3-qubit  PST-algorithm  for the $N=10$-node chain, $N^{(S)}_2= 3$, performed using 
the unitary transformation  $W_{ER}$ of  the particular form discussed  in Sec.\ref{Section:AWpart}. { Here $\lambda^2$ increases with $n^{(ER)}$ and increases with $Q$  for $n^{(ER)}=4,5$.  The fact that $\lambda^2 $ is almost the same for $Q=2$ and $3$ when $n^{(ER)}=6$ can be explained by the limitation of maximization accuracy  in the multiparametric case.} } 
\label{Table:T3}
\end{table}}

\end{document}